\PassOptionsToPackage{table}{xcolor}

\documentclass[11pt]{article}
\usepackage{jheppub}
\usepackage[normalem]{ulem}
\usepackage{physics}
\usepackage{slashed}
\usepackage{dsfont}
\usepackage{mathtools}
\usepackage{amsthm}
\usepackage{todonotes}
\usepackage{tikz,tikz-3dplot}
\usepackage{dsfont}
\usepackage{faktor}
\usepackage{pgfplots,pgfmath}
\usepackage{halloweenmath, mathrsfs}
\usepackage{multicol}
\usepackage{xfrac}
\usepackage{bm}
\usepackage{soul}

\usepackage{graphicx}
\graphicspath{ {./images} }

\usepackage{tikz,tikz-3dplot}
\usepackage{pgfplots}
\pgfplotsset{compat=1.17}
\usetikzlibrary{shapes,arrows,cd,chains,decorations.markings,decorations.pathmorphing,calc, patterns}
\tikzset{
	->-/.style args={#1rotate#2}{decoration={markings, mark=at position #1 with {\arrow[scale=1.5,rotate = #2 ]{stealth}}}, postaction={decorate}}
}
\tikzstyle{vertex}=[circle, draw, minimum size=0.2cm, inner sep = 1pt, fill=black] 
\tikzstyle{line}=[line width=0.3mm,->- = {0.5rotate0}]
\pgfmathsetmacro{\gS}{1}

\DeclareMathOperator{\bbR}{\mathbb{R}}

\DeclareMathOperator{\bbC}{\mathbb{C}}

\DeclareMathOperator{\calA}{\mathcal{A}}
\DeclareMathOperator{\calB}{\mathcal{B}}
\DeclareMathOperator{\calC}{\mathcal{C}}
\DeclareMathOperator{\calD}{\mathcal{D}}
\DeclareMathOperator{\calE}{\mathcal{E}}
\DeclareMathOperator{\calF}{\mathcal{F}}
\DeclareMathOperator{\calG}{\mathcal{G}}
\DeclareMathOperator{\calH}{\mathcal{H}}
\DeclareMathOperator{\calI}{\mathcal{I}}
\DeclareMathOperator{\calJ}{\mathcal{J}}
\DeclareMathOperator{\calM}{\mathcal{M}}
\DeclareMathOperator{\calN}{\mathcal{N}}
\DeclareMathOperator{\calP}{\mathcal{P}}
\DeclareMathOperator{\calS}{\mathcal{S}}
\DeclareMathOperator{\calQ}{\mathcal{Q}}

\DeclareMathOperator{\calO}{\mathcal{O}}
\DeclareMathOperator{\calL}{\mathcal{L}}

\DeclareMathOperator{\calY}{\mathcal{Y}}

\renewcommand{\dd}{\mathrm{d}}
\newcommand{\dvol}{\mathrm{dVol}}
\newcommand{\Qb}{\bar{Q}}
\newcommand{\calQb}{\bar{\mathcal{Q}}}
\newcommand{\Db}{\bar{D}}
\newcommand{\wt}[1]{\widetilde{#1}}

\DeclareMathOperator{\BV}{\mathrm{BV}}

\makeatletter
\DeclareRobustCommand{\extp}{%
  \mathchoice%
    {\extp@\displaystyle}%
    {\extp@\textstyle}%
    {\extp@\scriptstyle}%
    {\extp@\scriptscriptstyle}%
}
\newcommand{\extp@}[1]{%
  \mathop{\textstyle\Lambda}\nolimits^{\mspace{-6mu}#1}%
}
\makeatother

\newmuskip\pFqskip
\mathchardef\pFcomma=\mathcode`, 

\theoremstyle{definition}

\usepackage{xcolor}
\usepackage{booktabs, array, makecell, multirow}

\newcommand{\stoch}[1]{\expval{#1}_{\mathrm{st}}}

\usepackage{enumitem}

\setlist[itemize]{noitemsep}
\setlist[enumerate]{noitemsep}

\title{Dear Quantizers, Stochastic $\approx$ Brane}
\abstract{Stochastic quantization formulates $d$-dimensional Euclidean QFT as the equilibrium limit of a $(d+1)$-dimensional Langevin system. Using the supersymmetric formulation of stochastic systems, we apply techniques from Morse theory to study Langevin dynamics. We interpret stochastic quantization as the absolutization of a relative QFT by a bulk cohomological topological Equilibrium TFT (``EqmTFT''), and relate its enriched Neumann boundary data to the $d$-dimensional BV/BRST complex. Analytic continuation reveals connections to the A-model and non-relativistic and Carrollian limits.}

\author{Justin Kulp}

\affiliation{
Center for Cosmology and Particle Physics, 
    New York University, New York, NY 10003, USA\\
School of Natural Sciences, 
    Institute for Advanced Study, Princeton, NJ 08540, USA
}

\dedicated{For Saturn Grandma, with love from all cosmos creatures.}

\begin{document}
\maketitle

\section{Introduction and Summary}
Quantization of classical systems is a foundational problem in quantum mechanics and QFT. Textbook approaches to quantization typically invoke some form of ``canonical quantization,'' the promotion of classical phase space variables to quantum operators $f\mapsto \hat{f}$, replacing Poisson brackets with commutators $[\hat{f},\hat{g}] = -i\hbar\widehat{\{f,g\}}$; or ``path integral quantization,'' integrating over field configurations weighted by the classical action $e^{iS}$. After choosing one of these recipes, and an input classical theory, one argues that the output data satisfy \textit{some} axioms for quantum mechanics or QFT \cite{Dedushenko:2022zwd}. These quantizations are generally neither injective into nor surjective onto the space of QFTs.

Famously, both of these textbook approaches to quantization are also underdefined. Canonical quantization has the ``operator ordering problem'' (or ``Groenewold-Van Hove anomaly'') \cite{Groenewold:1946kp, van1951probleme} which states that the naive promotion of classical variables to quantum operators is ambiguous or impossible for general functions. This leads to more refined quantization schemes, including deformation quantization \cite{Bayen:1977ha, Fedosov:1994zz, Kontsevich:1997vb}, geometric quantization \cite{Souriau1966Quantification, kostant1972line, WoodhouseGeometric}, and brane quantization \cite{Gukov:2008ve, Gukov:2010sw, Gaiotto:2019oey} which require additional input to circumvent the underspecification. Likewise, path integral quantization involves an integral against a mathematically yet-to-be-defined measure and, even in finite-dimensional cases, the integrand may not be obviously convergent \cite{glimm2012quantum, guerra1975p} (see \cite{david2016liouville} for recent successes). A traditional strategy to avoid explicitly constructing the path integral is to make algebraic analogues, \`a la BV-BFV formalism \cite{Witten:1990wb, Henneaux:1992ig, gwilliam2012derive, johnson2015homological, Cattaneo:2019jpn, CG1, CG2}, or continue working at a physics level of rigour. Connections between different quantization schemes often highlight interesting properties of the output QFT and broadly connect different subjects in mathematical physics \cite{Cattaneo:1999fm, Cattaneo:2001bp, Kapustin:2001ij, Bressler:2002eu, Kapustin:2005vs, Witten:2010cx, Witten:2010zr, Kontsevich:2024esg}.

The stochastic quantization scheme of Parisi and Wu \cite{Parisi:1980ys} provides another approach to quantization based on stochastic time-evolution of classical fields.\footnote{Similar ideas were studied by Nelson \cite{Nelson:1966sp, nelson1973construction, nelson2020dynamical} and Symanzik \cite{Symanzik:1964zz,Jaffe:2014yka}.} Intuitively, if Euclidean QFT ``looks like'' statistical physics in equilibrium, then coupling a classical statistical system to a heat bath and waiting for it to equilibrate will make stochastic correlation functions resemble QFT correlation functions. In practice, the Parisi-Wu scheme promotes classical Euclidean fields to a function of time $\Phi(x) \mapsto \Phi(x,t)$ and defines time evolution by a Langevin equation
\begin{equation}\label{eq:LangevinIntro}
    \pdv{\Phi(x,t)}{t} = -\frac{1}{2}\fdv{S}{\Phi} + \eta(x,t)\,,
\end{equation}
with gradient potential and Gaussian white noise $\eta(x,t)$. The principal claim is then that stochastic correlation functions become QFT correlation functions at large times:
\begin{equation}
    \lim_{t\to\infty} \expval*{\Phi(x_1,t) \cdots \Phi(x_n,t)}_{\mathrm{st}} = \expval*{\Phi(x_1)\cdots\Phi(x_n)}_{\mathrm{QFT}}\,.\label{eq:StochCorr}
\end{equation}
The original physical motivation was to provide a quantization scheme for QFTs with gauge fields, while avoiding technical subtleties of gauge fixing conditions, by using a continuous time version of Monte Carlo. Recently, stochastic quantization has seen a resurgence of interest following a number of breakthroughs in stochastic partial differential equations and, subsequently, constructive QFT \cite{hairer2009introduction, hairer2014theory, gubinelli2021pde, chandra2022langevin, chandra2024stochastic} (see \cite{Jaffe:2014yka, gubinelliVideo} for helpful overviews).

Stochastic quantization raises a number of interesting and technical questions (even ignoring technical and analytic subtleties of constructive QFT):
\begin{enumerate}
    \item Which theories can be stochastically quantized? Can stochastic quantization accommodate fermions? What about Lorentzian theories? See \cite{Damgaard:1987rr} and references within for starters.
    \item Which stochastic process should we use for time-evolution? See \cite{parisi1982supersymmetric, Parisi:1983mgm, ovchinnikov2016introduction, Kaviraj:2019tbg, albeverio2019elliptic, Kaviraj:2020pwv, albeverio2020elliptic, Kaviraj:2021qii, gubinelliVideo, Rychkov:2023rgq, LeFloch:2025qjc, Sethi:2026xhb} for some alternatives to the Parisi-Wu scheme.
    \item Is the resulting theory reflection positive? See \cite{Jaffe:2014yka} for comments.
    \item Does the system actually equilibrate at large times? Is there a relationship between late-time correlation functions obtained this way and the usual textbook path integral quantization? See \cite{Cardy:1983aa, Gozzi:1983qxk, Gozzi:1984au, Damgaard:1987rr} for early work.
\end{enumerate}
The last problem is particularly interesting for us: while \eqref{eq:StochCorr} defines some correlation functions $\expval{\cdots}_{\mathrm{QFT}}$ which behave like correlation functions of a QFT, it is not immediately obvious that they are equivalent to standard path integral correlation functions. Schematically
\begin{equation}
    \expval{\Phi(x_1)\cdots \Phi(x_n)}_{\mathrm{QFT}} \stackrel{?}{=} \expval{\Phi(x_1)\cdots \Phi(x_n)}_{\mathrm{PI}}\,.\label{eq:equiv}
\end{equation}
A proof of perturbative equivalence was given by Parisi and Wu in their original paper \cite{Parisi:1980ys}, but non-perturbative equivalence is a little trickier. Some claims of non-perturbative equivalence in simplified setups are argued in loc. cit. and references within, modifying arguments by Cardy \cite{Cardy:1983aa}. One of our goals of this note is to discuss the equivalence in \eqref{eq:equiv} using arguments from supersymmetry.

One of the important and interesting features of the aforementioned canonical and path integral quantization schemes was that the fields and action were not sufficient data to define a QFT. For example, in geometric quantization, a symplectic manifold $(N,\omega)$ and prequantum line bundle $\mathfrak{L}$ do not specify a unique quantization; they require a polarization $\calP$. In brane quantization, an $A$-model and canonical coisotropic brane $\calB_{cc}$ do not specify a unique quantization; they require a choice of Lagrangian brane $\calB$. Finally, in path integral quantization, the path integral
\begin{equation}
    Z = \int [D\Phi] \, e^{-S[\Phi]}
\end{equation}
does not specify a theory, it requires a choice of integration cycle $\Gamma$ -- upon which $Z_{\Gamma}$ hopefully converges -- to define correlation functions. In slightly different ways, these examples show that $S$ provides information on how to formally/partially quantize an algebra of observables, consistent with the Dyson-Schwinger equations for $S$ (e.g., all $Z_{\Gamma}[J]$ correlation functions will satisfy the same Dyson-Schwinger equations), but additional choices are required to produce a correlation function or number. In other words, \textit{an action $S$ only specifies a relative quantum field theory}.

For simplicity, we will speak as if a choice of path integral cycle is all of the unspecified information of this relative theory in path integral quantization (i.e., we ignore potential global form subtleties, renormalization counterterms, etc.; these details are important but not the highlight of this note). Then the space of path integral cycles is the space of ``Dyson-Schwinger conformal blocks'' for theories with action $S$ \cite{Witten:2010zr, BVThimTFT}. Just like more familiar relative QFTs, e.g. WZW models \cite{Witten:1988hf, Elitzur:1989nr, Kapustin:2010if, Fuchs:2023ngi}, SymTFTs \cite{Kong_2017, Gaiotto:2020iye, Apruzzi:2021nmk}, or the 6d (2,0) theory \cite{Witten:1995zh, Witten:1998wy, Belov:2004ht, witten2004conformal,  Witten:2009at, Freed:2012bs, Gukov:2020btk}, there is a neat interpretation of these Dyson-Schwinger conformal blocks as states in a one-higher-dimensional theory \cite{Witten:2010cx, Witten:2010zr, BVThimTFT}: good cycles $\Gamma$ for 0d QFTs (integrals) are states of a 1d Supersymmetric Quantum Mechanics (SQM); good cycles $\Gamma$ for 1d QFTs (quantum mechanics) are states of an A-model; and so on. We leave further details to the main body of the paper and references. In any case, path integral correlation functions over the cycle $\Gamma$ take the form of a matrix element in the bulk theory
\begin{equation}
    \expval*{\calO(x_1)\cdots \calO(x_n)}_{\mathrm{PI},\Gamma}
        = \frac{1}{Z_{\Gamma}}\braket*{\Gamma}{\calO(x_1)\cdots\calO(x_n)}_{d+1}\,.\label{eq:thimbleOverlap}
\end{equation}

Returning to the stochastic quantization picture, a similar picture emerges essentially by definition: \textit{stochastic quantization defines a Euclidean QFT as a relative theory to a one-higher-dimensional bulk, with bulk dynamics determined by the action $S$, and absolutized by an equilibrium state.} Mathematically, we will argue that the stochastically quantized QFT correlation functions are a matrix element
\begin{equation}
    \expval*{\calO(x_1)\cdots \calO(x_n)}_{\mathrm{QFT}} = \frac{1}{Z_d}\braket*{\calM}{S;\calO(x_1)\cdots\calO(x_n)}_{d+1}\,.
\end{equation}
where $\bra*{\calM}$ is a special ``equilibrium out-state'' for the stochastic bulk. In this case, the ($d+1$)-dimensional bulk dynamics are described by a (generically) non-relativistic theory with an $\calN=2$ $D=1$ type supersymmetry, using the Parisi-Sourlas-Wu correspondence between stochastic PDEs and supersymmetric systems \cite{Nicolai:1980jc, Parisi:1980ys, parisi1982supersymmetric, cecotti1983stochastic, zinn2021quantum}.\footnote{These particular theories, and close cousins, were already studied in a variety of other contexts, including Conformal Quantum Critical Points (CQCPs) and RK states, Lifshitz supersymmetry, non-hermitian systems, Carrollian QFTs, and more (see e.g. \cite{Ardonne:2003wa, Dijkgraaf:2009gr, Chapman:2015wha, Masaoka:2025kiy, rokhsar1988superconductivity, Henley_2004, Castelnovo_2005, Jensen:2014wha, Henneaux:2021yzg, Cotler:2024xhb} and references within), giving a nice interpretation to some of their results as quantization procedures or bulk-boundary correspondences.} In the 0d/1d setup, this gives a direct match to the path integral cycle story in \eqref{eq:thimbleOverlap}, and it (seemingly) gives an alternative to the path integral cycle bulk in higher dimensions. This ($d+1$)-dimensional SQM reformulation also allows us to provide sharp geometric answers to the earlier questions about the space of states/equilibration and comment on theories involving fermions, gauge fields, and so on.

Perhaps the most essential feature of this supersymmetric reformulation of the Langevin dynamics is that we will typically care about $Q$-closed data of the ($d+1$)-dimensional SQM, modulo $Q$-exact data. Thus the SQM is really just scaffolding for a cohomological topological field theory (in the time direction), dubbed the ``Equilibrium TFT'' (``EqmTFT''), isolated by a topological twist. This allows us to understand the equivalence between stochastic quantization and path integral quantization as a localization argument, presented as a matrix element between states prepared at an ``enriched Neumann'' boundary and a state prepared at a ``reference'' boundary in the EqmTFT. This connects relative QFTs, and their absolutization, to derived QFT and quantization.

\paragraph{Outline of the Paper.} The paper is organized as follows:
\begin{enumerate}
    \item[{\hyperref[sec:Stochastic]{$\S$2.}}] Section \ref{sec:Stochastic} provides essential background connecting finite dimensional stochastic differential equations to 1d SQM, and contains a mixture of old and new results in a modern light. In Section \ref{sec:StochasticBackground} and Section \ref{sec:ParisiSourlasSUSY} we review textbook results on stochastic differential equations and their connection to supersymmetric systems. A key result is that coupled finite-dimensional gradient-type Langevin systems
    \begin{equation}
        \dot{q}^i(t) = -\frac{1}{2}\partial^i S(q) + \eta^i(t)\,,
    \end{equation}
    analogous to \eqref{eq:LangevinIntro} with fields $q^i$ on $M$ and 0d Euclidean action $S$, can be reformulated in terms of a 1d $\calN=2$ SQM with superfields $\Phi = (q,\psi,\bar\psi,B)$ on $M$ and superpotential $S$
    \begin{equation}
        S_{\mathrm{aux}}[\Phi] = \frac{1}{2\lambda}\int_{\calI}[\dd t|\dd\bar{\xi}\dd\xi](\Db\Phi^i D\Phi_i + S(\Phi))\,.
    \end{equation}
    In Section \ref{sec:SQMForBabies} we review the geometric interpretation of this SQM state space as a twisted de Rham complex.
    \vskip 0.125cm
    In Section \ref{sec:MorseTheory} we use the SQM formulation to understand the $0d$ partition function as a Rokhsar-Kivelson type state overlap
    \begin{equation}
        Z_{0d} = \int_M \dvol_M e^{-S(q)/\lambda} = \braket*{\Upsilon}\,,
    \end{equation}
    and decompose the state into Morse boundary states -- giving a real version of the familiar decomposition of partition functions into Lefschetz thimbles. In Section \ref{sec:BVNonsense} we use this to realize general 0d correlation functions as a state overlap
    \begin{equation}
    \expval*{\calO}_{\text{$0d$ QFT}} := \frac{1}{Z_{0d}} \braket*{\Upsilon}{S;\calO}
    \end{equation}
    between an enriched Neumann boundary state $\ket*{S;\calO}$ and a BPS (cohomological topological) reference boundary state $\bra*{\Upsilon}$. We give an interpretation of this overlap as a form of twisted Poincar\'e duality between BV cohomology classes and path integral cycles. In Section \ref{sec:NormalizabilityAnalyticCont} we review important normalizability conditions and sketch connections to Lefschetz thimbles and analytic continuation. In Section \ref{sec:FermionRant} we discuss SQM on supermanifolds.
    \vskip 0.25cm
    \item[{\hyperref[sec:higherDim]{$\S$3.}}] In Section \ref{sec:higherDim} we turn to the stochastic quantization of QFTs. In Section \ref{sec:StochasticGaussian} we modify our 0d/1d formulas to explain stochastic quantization and give the localization statement of its equivalence to usual path integral quantization. We also introduce the ``EqmTFT'' as the cohomological TFT underlying this localization procedure. In Section \ref{sec:InterpolationProof} we give a Cardy-like interpolating action argument, carefully recasting it using modern (worldsheet) supergeometry.
    \vskip 0.125cm
    In Example \ref{sec:AModel} we explicitly work through the stochastic quantization of quantum mechanics. We show that switching from a second-order to a first-order formulation of quantum mechanics forces us to make a choice of analytic continuation of the real Langevin equation. We show that a holomorphic extension leads to the complex Langevin equation of Parisi and Klauder, while a Hermitian extension leads to an A-model. In Example \ref{sec:WZWModel} we consider the CS/WZW correspondence. Instead of directly stochastically quantizing the theory, we show how identical manipulations can be used to realize the space of states of Chern-Simons theory as the BPS states in a Carrollian limit of supersymmetric Yang-Mills-Chern-Simons theory.
    \vskip 0.25cm
    \item[{\hyperref[app:SupersymmetryFormulas]{$\mathscr{A}$.}}]  In Appendix \ref{app:SupersymmetryFormulas} we review our SUSY conventions.
\end{enumerate}

\paragraph{Some Additional Thoughts.} Some comments and future directions include:
\begin{itemize}
    \item \textbf{More general Parisi-Sourlas-Wu maps.} The relationship between stochastic PDEs and supersymmetric systems is quite general. It would be interesting to understand how generic the stochastic PDEs can be and what their supersymmetric duals are (see also \cite{Sethi:2026xhb}). The manipulations should proceed quite similarly to this text.
    \vskip 0.125cm
    One particular example is the Parisi-Sourlas codimension-2 system \cite{parisi1982supersymmetric} (see \cite{Rychkov:2023rgq, LeFloch:2025qjc} for helpful reviews), whose stochastic interpretation relates some ($d+2$)-dimensional quenched disordered systems to $d$-dimensional systems. This case is particularly interesting because it involves non-nilpotent supercharges and equivariant localization computations, which could plausibly give relations to $\Omega$-deformations or the 4d/2d correspondence.
    \vskip 0.125cm
    However, one issue is that while we can use stochastic PDEs to generally inspire the study of a number of (non-relativistic) SQFTs, the map is not always exact. In cases like the Parisi-Sourlas codimension-2 system, signs in path integral Jacobians can change, leading to disagreements between the stochastic and supersymmetric formulations. An exact understanding of this issue warrants further consideration, as well as special attention paid to situations where it does not occur.
    \vskip 0.25cm
    \item \textbf{Formality of the bulk.} We discuss a formal procedure to quantize a $d$-dimensional theory by attaching it to a ($d+1$)-dimensional bulk. However, it is not obvious that the ($d+1$)-dimensional theory exists as a renormalizable and/or UV-complete QFT. For example, it is highly non-obvious that there is a renormalizable ($4+1$)d cohomological-TFT that describes the stochastic quantization of ($3+1$)d gauge theory.
    \vskip 0.125cm
    We speculate a few potential resolutions to this concern. First, it is completely possible that the ($d+1$)-dimensional theory exists in low dimensions, but only formally in general dimensions. Barring that, one speculation is that a bulk always exists as an interacting \textit{non-relativistic} fixed point. Indeed, general Lifshitz powercounting naturally pushes up critical dimensions, e.g., the $z=2$ analog of the Wilson-Fisher fixed point lives in ($3+1$)d \cite{Nishida:2007pj}, so that interacting renormalizable non-relativistic theories could exist in higher dimensions.\footnote{There are many examples where intuitions about the space of unitary QFTs are violated by dropping assumptions of Lorentz invariance. E.g., one can flow between relativistic CFTs with $c_{\mathrm{IR}} > c_{\mathrm{UV}}$ by leaving the space of relativistic QFTs \cite{Ammon:2011nk}. Generally, the space of non-relativistic fixed points seems quite unexplored.} Alternatively, even if the mixed-topological bulk theory is non-renormalizable and requires an infinite number of counterterms, it is possible that these infinitely many counterterms could be chosen in an essentially unique way. This is reminiscent of what happens in BCOV theory \cite{Costello:2015xsa}, and can be understood as the statement that BCOV theory has a unique (perturbative) quantization.
    \vskip 0.25cm
    \item \textbf{Analytic continuations.} At many points throughout the text, we find ourselves naturally wanting to analytically continue the real Langevin equation. This happens when we consider complex actions like first-order Euclidean theories or the WZW model. This first point is particularly interesting because quantizing a first-order theory should lead to a relativistic bulk. Regardless of motivation, one must choose the specific equations they are analytically continuing -- and the choice of continuation -- very carefully, because analytically continuing at different stages leads to different results. It seems very important to understand these choices and the relations between them. We speculate without further justification that some relations between complex Langevin, AKSZ/Poisson sigma models \cite{Alexandrov:1995kv, Khan:2025rah, BVThimTFT}, and the analytic continuation of the path integral will emerge with careful treatment.
\end{itemize}

\section{Stochastic Differential Equations and Supersymmetry in One Dimension}\label{sec:Stochastic}
In this section, we focus entirely on the Langevin equation on finite-dimensional manifolds, relevant to stochastic quantization of 0d QFTs (integrals) by a 1d SQM, where everything is explicitly computable. As we will see, extension to higher dimensions in Section \ref{sec:higherDim} becomes essentially trivial once we have organized everything correctly in this 0d/1d case.

In Section \ref{sec:StochasticBackground} we review the Langevin and Fokker-Planck equations, with an eye towards their equilibrium distributions and the special role of gradient potentials. In Section \ref{sec:ParisiSourlasSUSY} we give a path integral interpretation to the Langevin and Fokker-Planck equations, and review the well-known connection between stochastic differential equations and supersymmetric systems \cite{Nicolai:1980jc, Parisi:1980ys, parisi1982supersymmetric, cecotti1983stochastic, zinn2021quantum}. Emphasis is placed on the treatment of fermionic determinants and the validity of the supersymmetric reformulation. Perhaps more importantly, we give consistent conventions for SUSY fields and SUSY transformations (see also Appendix \ref{app:SupersymmetryFormulas}), which we use in our lightning review of essential facts of $\calN=2$ SQM in Section \ref{sec:SQMForBabies}.

In Section \ref{sec:MorseTheory} we apply results from SQM to the Langevin dynamics. We pay particular attention to the normalizability of states and/or existence of an equilibrium state, the natural relation of the equilibration process to Morse flows and Morse boundary conditions, and show how 0d correlation functions emerge from localization. In Section \ref{sec:BVNonsense} we give a new interpretation of the Langevin process in terms of boundary BV data and explain the equilibration process as a form of absolutization of a relative QFT specified by this boundary BV data. This gives tantalizing hints towards a non-perturbative BV formalism using the higher-dimensional bulk. In Section \ref{sec:NormalizabilityAnalyticCont} we expound further upon issues of normalizability, the relationship to analytic continuation, and the role played by different cycles for the path integral as a form of twisted Poincar\'e duality. In Section \ref{sec:FermionRant} we comment on extensions to fermionic theories.

\subsection{Survey of the Langevin and Fokker-Planck Equations}\label{sec:StochasticBackground}
To represent systems with random fluctuations, we make use of classical stochastic variables and stochastic DEs. In this section, we will state some elementary definitions and facts about stochastic variables at a physics level of rigour \cite{Damgaard:1987rr, zinn2021quantum}.\footnote{We warn the reader that physics references sometimes collapse fundamental definitions in a confusing way. We highly recommend the interested reader read Section 2 of \cite{hairerMarkov} and \cite{hairer2009introduction} for precise definitions. As a quick translation guide, a random variable is just a hom of measurable spaces (aka ``measurable map'') $X:(\Omega,\calF) \to (\mathcal{X},\calA)$ where the sample space $(\Omega,\calF)$ is equipped with a probability measure $\mathbb{P}$. The target $(\mathcal{X},\calA)$ is called the state space. Crucially, the random variable $X$ pushes forward $\mathbb{P}$ to a probability measure $\mu_X := X_{*}\mathbb{P}$ on $(\mathcal{X},\calA)$. Physics references sometimes only use this induced probability measure and also use the terms state space and sample space interchangeably. E.g., for a real random variable $X$, we take $(\mathcal{X},\calA) = (\bbR, \mathcal{B}(\bbR))$, and the probability that the random-variable $X$ takes values in a set $A$ is $\mu_X(A)$. Below we also assume $\mu_X$ admits a probability distribution $\mu_X(A) =: \int_A P_X(x) \, \dd x$.} Then we will see that a close relation to (pseudo-Hermitian, SUSY) quantum mechanics emerges.

A ``stochastic variable'' (aka ``random variable'') $X$ is a variable taking values in a ``state space'' $\mathcal{X}$ (which we treat here as if it is $\bbR^m$ for simplicity). Since this is classical probability theory, a probability distribution is a normalized, non-negative, real function $P_X(x)$, and the probability that $X \in A \subseteq \mathcal{X}$ is
\begin{equation}
    P(X \in A) = \int_{A} \dd^m x\, P_X(x)\,.
\end{equation}
Expectation values of functions of random variables are given by
\begin{equation}
    \expval{f(X)}_{\mathrm{st}} := \int_{\mathcal{X}} \dd^mx\, f(x) P_X(x)\,.
\end{equation}
We introduce $\stoch{\cdots}$ to emphasize that this is a classical statistical average.
 
A ``stochastic process'' $Y$ is a \textit{collection} of random variables $\{Y_t\}_{t\in I}$ indexed by a time parameter $t$. The random variable $Y_t$ is called the ``value of the stochastic process at time $t$,'' and we assume all $Y_t$ share a common state space $\mathcal{Y}$ (which we treat here as if it is $\bbR^n$ for simplicity). Thus, for each fixed $t$, there is a random variable $Y_t$ and an associated distribution $P_{Y_t}(y)$ against which we can compute expectation values of functions $f(Y_t)$, etc. We can also consider expectation values of functions of two random variables at different times
\begin{equation}
    \expval{f(Y_{t_1}, Y_{t_2})}_{\mathrm{st}} = \int_{\calY\times\calY} \!\!\!\!\!\!\dd^n y_1 \dd^n y_2 \, f(y_1, y_2) P_{t_1, t_2}(y_1, y_2)\,,
\end{equation}
where $P_{t_1, t_2}(y_1,y_2)$ is the joint probability distribution, and so on. 

A particularly nice way to define a stochastic process $Y$ is to set $Y_t := g(t,X)$ for some random variable $X$ and function $g:I\times \mathcal{X} \to \mathcal{Y}$. Then the expectation value of functions of $Y_t$ is inherited from $X$ by composition,
\begin{equation}\label{eq:compositionProbability}
    \expval{f(Y_t)}_{\mathrm{st}} = \int_{\mathcal{X}} \dd^m x\, f(g(t,x)) P_X(x)\,,
\end{equation}
as are joint probability distributions
\begin{equation}
    P_{t_1,\dots, t_\ell}(y_1,\dots, y_\ell) := \int_{\mathcal{X}} \dd^mx\, \delta^{(n)}(y_1 - g(t_1, x)) \cdots \delta^{(n)}(y_\ell - g(t_\ell, x)) P_X(x)\,.
\end{equation}
In this nice case, it is useful to recast the entire stochastic process $Y$ as a single random variable with target state space given by the trajectories $\calM := \mathrm{Maps}(I,\mathcal{Y})$. Define $G:\mathcal{X}\to \calM$ by $G(x) := g(\cdot, x)$, then we can define expectation values of $Y$ on the space of trajectories
\begin{equation}
    \expval{F[Y]}_{\mathrm{st}} = \int_{\mathcal{M}} [\dd y]\, F[y] P_Y[y]\,.
\end{equation}
$P_Y[y]$ is again inherited from $P_X(x)$ by composition
\begin{equation}
    P_Y[y] := \int_{\mathcal{X}} \dd^m x\, \delta[y-G(x)] P_X(x)\,.
\end{equation}
The measure $[\dd y]$ is just a product of $[\dd^n y_i]$ for finite sets $I$, and we leave the infinite case to the formal machinery of the references. This repackaging in terms of trajectories contains all the information from before, e.g., the joint probability distribution is
\begin{equation}
    P_{t_1, t_2}(y_1, y_2) = \int_{\mathcal{M}} [\dd y]\, \delta^{(n)}(y_1 - y(t_1))\delta^{(n)}(y_2 - y(t_2)) P_Y[y]\,.
\end{equation}
I.e., it is the tiny slice of trajectories $y$ that pass through $y_1$ at $t_1$ and $y_2$ at $t_2$, weighted by the trajectory probability density $P_Y[y]$.

Most interesting for us are the ``Markov stochastic processes.'' For simplicity, assume that $I$ is a discrete set of times $t_1,t_2,\dots, t_\ell, \dots$. A stochastic process is called Markov if the conditional probabilities satisfy
\begin{equation}
    P(y_{\ell+1},t_{\ell+1}|y_{1},t_{1}; \dots ; y_{\ell},t_{\ell}) = P(y_{\ell+1},t_{\ell+1}|y_{\ell},t_{\ell})\,.
\end{equation}
In other words, ``knowing the entire history of the process does not contain any more information than knowing its last value'' (see \cite{hairerMarkov} up to and including Section 2.2 for rigorous details). The continuous time analog likewise says that the conditional probability $Y(t) = y$ only depends on the state of the system at $t - \delta t$. 

In our case, we are interested in the relation between two stochastic processes $q$ and $\eta$ connected by a Stochastic Differential Equation (SDE): a differential equation involving trajectory random variables.\footnote{The formal tools of stochastic calculus require some care in their definition, and can be found in \cite{Damgaard:1987rr, zinn2021quantum, hairer2009introduction}. We are using the ``Stratonovich stochastic calculus'' because it allows formal manipulations analogous to usual differential calculus. To the best of our understanding, it is equivalent to the more proof-useful ``It\^o calculus.'' Likewise, we follow physics references \cite{Damgaard:1987rr, zinn2021quantum} in treating our Gaussian white-noise like a generalized free field, rather than formally introducing the Wiener process $W_t = \int_0^t \dd\tau\, \eta$ and writing S(P)DEs in terms of $\dd W_t$.} Specifically, we are interested in the family of first-order SDEs taking the form of a \textit{Langevin equation}:
\begin{equation}\label{eq:LangevinEquation}
    \dot{q}^i (t) = -\frac{1}{2} f^i(q(t)) + \eta^i(t)\,.
\end{equation}
Here, $q: I \to \bbR^n$ is a trajectory in $\bbR^n$ (for simplicity) and $\eta^i$ is ``Gaussian white noise.'' The Gaussian white noise satisfies Wick-like rules: the one-point function is vanishing, the two-point function is defined to be
\begin{equation}
    \expval*{\eta^i(t)\eta^j(t')}_{\mathrm{st}} := \lambda \delta^{ij} \delta(t-t')
\end{equation}
for some $\lambda$, and other correlators are defined/computed by Wick-contractions. This can also be computed by a functional integral against:
\begin{equation}
    \frac{e^{-\frac{1}{2\lambda} \int \dd t \, \eta^2(t)}}{\int [\dd\eta]\, e^{-\frac{1}{2\lambda} \int \dd t \, \eta^2(t)}}\, = C e^{-\frac{1}{2\lambda} \int \dd t \, \eta^2(t)}\,.
\end{equation}
The statement that the noise is white refers to the fact that it is completely uncorrelated from instant to instant in time (i.e., it is even stronger than Markov). 

If we treat the stochastic process $\eta$ as a trajectory random variable, then solving the SDE gives a map $q_t = g(t,\eta; q_0, t_0)$ defining the stochastic process $q$ based on the trajectory random variable $\eta$ and initial data. In practice, this means we explicitly solve the differential equation \eqref{eq:LangevinEquation} for the trajectory $q$ with initial conditions $q(t_0) = q_0$ and fixed noise-configuration, then statistically average over all noise trajectories to obtain expectation values of $q$ as in \eqref{eq:compositionProbability} (see Example \ref{sec:BrownianExample}). Thanks to the time-local structure of $f^i$, the solution $q$ is necessarily a Markov stochastic process, giving a continuous time analog of a Monte Carlo Markov chain.

Since the process is Markovian, the conditional law of $q$ at time $t$ only depends on the value of the trajectory $q$ at time $t-\delta t$. Extrapolating backwards in time, we can compute the conditional probability $P(\tilde{q},t;q_0,t_0)$ that the random variable $q_t$ equals $\tilde{q}$ given the initial condition $q(t_0)=q_0$. Thus we can write expectation values of functions of $q_t$ in terms of the initial data
\begin{equation}\label{eq:integrateOutAndIn}
    \stoch{F(q_t)} = C \int [\dd\eta]\, F(q_t)\,  e^{-\frac{1}{2\lambda} \int \dd t\, \eta^2(t)} = \int_{\bbR^n} \dd \tilde{q} \, F(\tilde{q}) P(\tilde{q},t;q_0,t_0)\,,
\end{equation}
where $P(\tilde{q},t;q_0,t_0) = \expval*{\delta^{(n)}(q_t-\tilde{q})}_{\mathrm{st}}$. Here we have turned the functional integral over the trajectory random variable $\eta$ on the left into a genuine finite-dimensional integral over configurations $\tilde{q} \in \bbR^n$ at time $t$ on the right. Going forward, we will switch to physics notation and write $q(t) = q_t$, use $q(t)$ to talk about both the trajectory and its evaluation at a particular time, and drop $\sim$ on $\tilde{q}$.

\paragraph{The Fokker-Planck Equation.} The conditional probability satisfies a number of simple properties (see \cite{zinn2021quantum}, Ch. 4), including locality and a semi-group composition in time. Thus it is useful to view the conditional probability as the kernel of a time-evolution operator $P$, and introduce ``bras'' and ``kets'' so that:
\begin{equation}
    P(q,t;q_0,t_0) =: \mel*{q}{P(t;t_0)}{q_0}\,.
\end{equation}
We further define the (non-Hermitian)\footnote{In general, $P(t;t_0)$ is a real kernel and thus the eigenvalues of $H_{\mathrm{FP}}$ are real or complex conjugate pairs. If we assume $H_{\mathrm{FP}}$ is diagonalizable with discrete spectrum, then it is ``pseudo-Hermitian'' with respect to some $\zeta$, meaning there exists a Hermitian $\zeta$ such that $H_{\mathrm{FP}}^\dagger = \zeta H_{\mathrm{FP}} \zeta^{-1}$. See \cite{Mostafazadeh:2002sk, ovchinnikov2016introduction} for generalities and the (non-unique) construction of $\zeta$. Pseudo-hermiticity is equivalent to ``$\zeta$T-symmetry'' in finite-dimensional systems \cite{Bender:1998ke, Mostafazadeh:2001jk,Mostafazadeh:2002sk, Mostafazadeh:2001nr, Bender:2007nj, Mostafazadeh:2008pw}. Assuming diagonalizability, the left and right eigenvectors, $\bra{\psi_{L,n}}$ and $\ket{\psi_{R,n}}$, of $H_{\mathrm{FP}}$ form a complete biorthonormal system, i.e. $\mathds{1} = \sum_n \ket{\psi_{R,n}}\!\!\bra{\psi_{L,n}}$. Soon we will turn to Hermitian Hamiltonians arising in detailed balance systems, but many of the structures that we will describe, such as supersymmetry and $Q$-exactness, still persist with appropriate pseudo-modifications (see again \cite{Mostafazadeh:2002sk, Ovchinnikov:2012pb, Ovchinnikov:2012qa, ovchinnikov2016introduction, Ovchinnikov:2026jvd, Sethi:2026xhb}).} Fokker-Planck Hamiltonian $H_{\mathrm{FP}}$ by:
\begin{equation}
    P(t;t_0) = e^{-(t-t_0)H_{\mathrm{FP}}}\,.
\end{equation}
A straightforward computation shows that \cite{zinn2021quantum}:
\begin{align}\label{eq:FokkerPlanckHamiltonian}
        H_{\mathrm{FP}} 
        &= -\frac{\lambda}{2}\partial_i(\partial^i + \tfrac{1}{\lambda}f^i(\hat{q}))\,.
\end{align}
The Markov property of the conditional probability distribution is captured by the \textit{Fokker-Planck equation}:
\begin{equation}\label{eq:FokkerPlanck}
    \partial_t P(t;t_0) = - H_{\mathrm{FP}} P(t;t_0)\,.
\end{equation}
It is standard to suppress the initial conditions, and we will do so in the remainder of the discussion when no confusion should arise.

At this point, the analogy to quantum mechanics is clear: we replace ``quantum fluctuations'' by our stochastic (P)DE coupled to random noise. The strength of noise correlations $\lambda$ is analogous to $\hbar$. The probability distribution $P(q,t)$ is the transition amplitude $\braket{q,t}{q_0,t_0} = \bra{q}\!e^{-(t-t_0)H_{\mathrm{FP}}}\!\ket{q_0}$, telling us the probability that we are in the state $q$ at time $t$ given our initial conditions, and the Fokker-Planck equation is our analog of the Euclidean Schr\"odinger equation.

We could also generalize the previous situation, allowing general maps $q: I \to M$ into some target manifold $(M,g)$, and adding various curvature terms to the derivations above. We will present the case $M=\bbR^n$ for notational simplicity, as general $M$ lead to a number of cluttering curvature terms, obscuring the analysis.

\paragraph{Equilibrium Distributions.} In favourable situations, it is conceivable that the stochastic system tends to an (unnormalized) \textit{equilibrium distribution}:
\begin{equation}\label{eq:eqmDist}
    P_{\mathrm{eq}}(q) := Z_{0d} \lim_{t\to\infty} P(q,t) =: e^{-S(q)/\lambda}\,.
\end{equation}
By definition, the equilibrium distribution is a time-stable configuration, so \eqref{eq:FokkerPlanck} implies that $P_{\mathrm{eq}}(q)$ is a right eigenfunction of $H_{\mathrm{FP}}$ with eigenvalue $0$. By conservation of probability, it can be argued that a left eigenfunction with eigenvalue $0$ is a constant in $q$ \cite{zinn2021quantum}. Thus we introduce the left-eigenvector $\bra*{0_{\mathrm{L}}}$ and right-eigenvector $\ket{\mathrm{eq}}$ of $H_{\mathrm{FP}}$ with eigenvalue $0$, which then satisfy $\braket*{0_\mathrm{L}}{q} = 1$, $\braket{q}{\mathrm{eq}} = P_{\mathrm{eq}}(q)$ and
\begin{equation}\label{eq:GroundStateProbability}
    Z_{0d} := \braket*{0_\mathrm{L}}{\mathrm{eq}} = \int_{\bbR^n} \dd^n q\,P_\mathrm{eq}(q) < \infty\,.
\end{equation}
$H_{\mathrm{FP}}$ can be gapped or gapless and $\ket{\mathrm{eq}}$ is the ground state of $H_{\mathrm{FP}}$ if it exists. This is compatible with the analogy to Euclidean quantum mechanics: the $t\to\infty$ limit gives the ground state.

If we assume an equilibrium configuration exists, then it constrains the form of the Langevin equation and Fokker-Planck Hamiltonian \cite{zinn2021quantum}. In particular, if $e^{-S(q)/\lambda}$ is going to be an equilibrium distribution, then $f^i$ must have the form
\begin{equation}\label{eq:GSDetermined}
    f^i(q) = \partial^i S(q) + V^i(q) e^{S(q)/\lambda}\,,
\end{equation}
for some $\partial_i V^i(q) = 0$. Note, the converse is not true, not all $f^i(q)$ of this form admit an equilibrium distribution.

As mentioned, there is no guarantee that the limit in \eqref{eq:eqmDist} actually exists. Physically, an equilibrium distribution could not exist if the Langevin equation has ``runaway solutions,'' where the probability of finding $q(t)$ inside a large finite radius ball goes to zero as $t\to\infty$. In this case, $0$ is not actually in the normalizable spectrum of $H_{\mathrm{FP}}$ and $\ket{\mathrm{eq}}$ is not actually in the state space. For example, when $q$ has a non-compact target like $\bbR^n$, the inner-product $\braket*{0_\mathrm{L}}{\mathrm{eq}}$ may depend sensitively on the parameters of the problem, as we see in Example \ref{sec:BrownianExample}. There could also plausibly be multiple states with $0$ eigenvalue. This follows more cleanly from the supersymmetric formulation and the discussion in Sections \ref{sec:MorseTheory} and \ref{sec:NormalizabilityAnalyticCont}.

\paragraph{Detailed Balance and Unitarity.} Since $H_{\mathrm{FP}}$ was not Hermitian, but only pseudo-Hermitian, there is no sense in which the equilibrium distribution in \eqref{eq:eqmDist} integrates to the norm of one vector called $\ket{\mathrm{eq}}$, i.e., $\bra*{0_\mathrm{L}}$ is not the Hermitian conjugate to $\ket{\mathrm{eq}}$. However, a Hermitian operator \textit{can} be defined in the special case of a reversible Langevin equation, or ``detailed balance'' process, i.e. when the Langevin equation is a gradient flow coupled to white noise:
\begin{equation}\label{eq:gradientFlow}
    f^i(q) = \partial^i S(q)\,.
\end{equation}
In such cases, we are motivated to define a new evolution operator
\begin{equation}\label{eq:newEvolution}
    \mel*{q}{U(t;t_0)}{q_0} := e^{S(q)/2\lambda} \mel*{q}{P(t;t_0)}{q_0} e^{-S(q_0)/2\lambda}\,,
\end{equation}
which satisfies the corresponding Euclidean Schr\"odinger equation
\begin{equation}
    \partial_t U(t;t_0) = - H U(t;t_0)\,,
\end{equation}
with Hermitian Hamiltonian $H$ given by
\begin{equation}\label{eq:susyH}
    H = \frac{\lambda}{2} \hat{D}_i^\dagger \hat{D}^i\,,
\end{equation}
where
\begin{align}
\begin{split}\label{eq:covariantDerivative}
    \hat{D}_i := e^{S/2\lambda} \, (\partial_i + \tfrac{1}{\lambda}\partial_iS) \, e^{-S/2\lambda} = \partial_i + \frac{1}{2\lambda} \partial_i S\,,\\
    \hat{D}^\dagger_i := e^{S/2\lambda} \, (-\partial_i) \, e^{-S/2\lambda} = -\partial_i + \frac{1}{2\lambda} \partial_i S\,.
\end{split}
\end{align}
Since $H$ is related to $H_{\mathrm{FP}}$ by a similarity transformation, the two formally have the same spectrum. In particular, when the $0$ eigenvalue is actually in the normalizable spectrum of $H$, the left and right eigenvectors $\bra{0}$ and $\ket{0}$ corresponding to the $0$ eigenvalue of $H$ are actually the same $L^2$-normalizable ground state, with $\braket{q}{0} = e^{-S(q)/2\lambda}$. 


With foresight, we note that the Hamiltonian \eqref{eq:susyH} could be viewed as a component Hamiltonian in a supersymmetric quantum mechanics. If the Hamiltonian
\begin{equation}\label{eq:Hplus}
    H_+ := H =  
        \frac{\lambda}{2}\left(-\partial^2 
        - \frac{1}{2\lambda}\partial^2S 
        + \frac{1}{4\lambda^2}(\partial_i S)^2 \right)\,,
\end{equation}
corresponds to the component with fermion number $n$, then the Hamiltonian in the fermion number $0$ sector is
\begin{equation}\label{eq:Hminus}
    H_{-} 
    := \frac{\lambda}{2}\hat{D}_i \hat{D}^{i\dagger}
    = \frac{\lambda}{2}\left(-\partial^2 
        + \frac{1}{2\lambda}\partial^2S 
        + \frac{1}{4\lambda^2}(\partial_i S)^2 \right)\,.
\end{equation}
This new $H_-$ is conjugate to the Fokker-Planck Hamiltonian with reversed gradient flow potential $S \mapsto -S$.

\subsubsection{Example: Brownian Motion of a Particle}\label{sec:BrownianExample}
The \textit{Brownian motion} of a particle in $\bbR^n$ is described by a linear Langevin equation\footnote{In the notation of the previous section $q^i = m v^i$ and $f^i(q) = \frac{2\alpha}{m} q^i = \frac{\alpha}{m}\partial^i q^2$, with dissipation factor $\alpha$.}
\begin{equation}
    m \frac{d}{dt} \vec{v}(t) = - \alpha \vec{v}(t) + \vec{\eta}(t)\,.
\end{equation}
Langevin's equation models both the deterministic dissipative effects of a fluid on a particle plus changes to the velocity from random buffeting by external molecules. Note that each classical velocity $v_i(t)$ in the Langevin equation is coupled to an independent stochastic white noise variable $\eta_i(t)$. This is what we would expect: the random thermal fluctuations in different directions should be independent.

We can give a formal solution to the Langevin equation with initial condition $v(t_0) = v_0$ and some fixed noise-configuration:
\begin{equation}\label{eq:vSoln}
    v(t) = v_0 e^{-\frac{\alpha}{m}(t-t_0)} + \frac{1}{m}\int_{t_0}^t d\tau\, e^{-\frac{\alpha}{m}(t-\tau)} \eta(\tau)\,.
\end{equation}
In principle, we can now compute the stochastic expectation value of functions of $v(t)$ in terms of $\eta(t)$. For example, suppose $v(0) = 0 $ and $0 < t \leq t'$, then
\begin{equation}
    \frac{m}{2}\expval*{v_i(t) v_j(t')}_{\mathrm{st}} = \delta_{ij} \frac{\lambda}{2\alpha}\sinh\left(\frac{\alpha}{m}t\right)e^{-\frac{\alpha}{m}t'}\,.
\end{equation}
Note that, after long times, the average kinetic energy becomes
\begin{equation}
    E_{\mathrm{avg}}^{\mathrm{th}} = \lim_{t\to\infty} \frac{m}{2}\stoch{v_i(t) v^i(t)} = \frac{n}{4}\frac{\lambda}{\alpha}
\end{equation}
if $\alpha > 0$. Intuitively, the Langevin equation describes a system that reaches a steady state, or thermal equilibrium, after long times, when there is a damping force. When $\alpha < 0$ the system does not reach an equilibrium distribution because the potential is not confining.

Instead of working with products of \eqref{eq:vSoln} and performing Wick contractions of $\eta$'s in $\stoch{\cdots}$, we can use the Fokker-Planck formulation. Indeed, since the force is a gradient, we can write the Hermitian Hamiltonian
\begin{equation}
    H_+ = -\frac{\lambda}{2m^2} \frac{\partial^2}{\partial v^2} + \frac{\alpha^2}{2\lambda} v^2 - \frac{n \alpha}{2m}\,.
\end{equation}
This is just a harmonic oscillator (be warned that $v$ is a coordinate here), and the ground state is:
\begin{equation}
    \psi_0(v) = \exp\left(-\frac{m\alpha}{2 \lambda} v^2\right)\,.
\end{equation}
The (properly normalized) equilibrium distribution is just the Maxwell-Boltzmann distribution for the Brownian particle:
\begin{equation}
    P_{\mathrm{eq}}(v) = \left(\frac{m \alpha}{\pi\lambda}\right)^{\tfrac{n}{2}}\exp\left(-\frac{m\alpha}{\lambda} v^2\right)\,,
\end{equation}
and we can easily recover the average energy at long times from the expectation value of $v^2$ in the ground state:
\begin{equation}
    E^{\mathrm{th}}_{\mathrm{avg}} = \frac{m}{2}\int_{\bbR^n} \dd v\, v^2 P_{\mathrm{eq}}(v) = \frac{n}{4}\frac{\lambda}{\alpha}\,.
\end{equation}

\subsection{Path Integrals and Parisi-Sourlas-Wu Supersymmetry}\label{sec:ParisiSourlasSUSY}
Now we turn to a path integral formulation of the previous discussion and review the supersymmetric formulation of the problem. 

We start by bringing the stochastic path integral,
\begin{equation}\label{eq:stochPI}
    Z_{\mathrm{st}}[J] :=  \int [\dd\eta]\, e^{-\frac{1}{2\lambda} \int \dd t\, \eta^2 - \int \dd t\, J q}\,,
\end{equation}
to Fokker-Planck form. As in \eqref{eq:integrateOutAndIn}, we integrate in $q$ and integrate out $\eta$, but allowing for arbitrary insertions in time.\footnote{As in elementary quantum mechanics, we can also obtain a path integral for the Fokker-Planck formulation by inserting many factors of $\int \dd q_i \ket{q_i}\!\!\bra{q_i}$ in transition amplitudes $\braket{q,t}{q_0,t_0} = \mel{q}{P(t;t_0)}{q_0}$.} Following \cite{Gozzi:1983qxk, Damgaard:1987rr, zinn2021quantum}, we define
\begin{equation}
    E^i(q(t),\eta(t)) := \dot{q}^i(t) + \frac{1}{2}f^i(q(t)) - \eta^i(t)\,,
\end{equation}
so that the Langevin equation \eqref{eq:LangevinEquation} is just $E^i(q(t),\eta(t)) = 0$. We also define
\begin{equation}
    \Delta^i_j(t,t') := \frac{\delta E^i(t)}{\delta q^j(t')} = \left[\delta^i_{j}\partial_t + \frac{1}{2}\frac{\partial f^i}{\partial q^j}\right]\delta(t-t')\,.
\end{equation}
The determinant of $\Delta$ can be factored as $\det \Delta = \det(\partial_t) \det(\Delta_{\pm})$ where
\begin{equation}
    \Delta_{\pm,j}^i(t,t') := \delta^i_j \delta(t-t') \pm \Theta(\pm(t-t'))\frac{1}{2}\partial_j f^i\,.
\end{equation}
The different choices $\Delta_{\pm}$ reflect the non-uniqueness of the Green's function $\partial_t^{-1}$; obviously convex linear combinations will work as well. Forward and backward-time Langevin evolution are given by $\Delta_+$ and $\Delta_-$ respectively. A textbook calculation gives
\begin{equation}
    \det \Delta_{\pm} = \exp \tr \log \Delta_{\pm} = \exp(\pm\frac{1}{4}\int_I \dd t'\, \partial_i f^i)\,.
\end{equation}

Now we can recast the stochastic path integral \eqref{eq:stochPI} in the Fokker-Planck form:
\begin{align}
    Z_{\mathrm{FP},\pm}[J]    
        &:= \int [\dd\eta] [\dd q]\, \det\Delta_{\pm} \, \delta(E^i(q,\eta))\, e^{-\frac{1}{2\lambda} \int \dd t\, \eta^2 - \int \dd t\, J q}\label{eq:PIqeta}\\
        &= \int [\dd q]\, e^{-\frac{1}{2\lambda}\int \dd t\, \left[(\dot{q}+\frac{1}{2}f(q))^2 \mp \frac{\lambda}{2} \partial_i f^i\right] - \int \dd t\, J q}\label{eq:PIjustq}
\end{align}
We will call the action in \eqref{eq:PIjustq} the Fokker-Planck action $S_{\mathrm{FP},\pm}$. We have ignored the $\det(\partial_t)$ term since it will simply shift the path integral by terms which will cancel in normalized correlation functions.

When we have a detailed-balance force $f^i(q) = \partial^i S(q)$, the Fokker-Planck action can be simplified by identifying a total derivative in $t$ and integrating:
\begin{align}
    S_{\mathrm{FP},\pm} 
        &= \frac{1}{2\lambda}\int_I \dd t\, \left[\left(\dot{q}_i+ \tfrac{1}{2}\partial_i S\right)^2 \mp \tfrac{\lambda}{2}\partial^2S\right]\\
        &= \frac{1}{2\lambda} \int_I \dd t\, \left[\dot{q}_i^2 \mp \tfrac{\lambda}{2} \partial^2 S + \tfrac{1}{4}(\partial_i S)^2\right] + \frac{1}{2\lambda} (S(q_f) - S(q_i))\\
        &=: S_{\pm} + \tfrac{1}{2\lambda}S(q)\big\vert_{q_i}^{q_f}\,.
\end{align}
The new $S_{\pm}$ is the action for the Hamiltonians $H_{\pm}$ in \eqref{eq:Hplus} and \eqref{eq:Hminus}.\footnote{It may seem surprising that the difference between the Fokker-Planck action $S_{\mathrm{FP},\pm}$ and unitary action $S_{\pm}$ is given by a boundary term. Generally, if we have any Hamiltonian $\hat{H}$ and conjugate it by a position-dependent piece $\hat{H} \mapsto e^{V(\hat{q})}\hat{H} e^{-V(\hat{q})}$ (as in \eqref{eq:newEvolution}), then the action changes by the endpoint differences $S \mapsto S - [V(q_f) - V(q_i)]$. Indeed, if we view the 1d quantum mechanics like the worldvolume of a Wilson line, the conjugation is a $U(1)$ gauge transformation which accumulates at the endpoints. We saw this in \eqref{eq:covariantDerivative} when the derivative $\partial_i$ ``in Fokker-Planck gauge'' was transformed to a covariant derivative ``in the unitary gauge.'' Thus, the non-trivial observation is really that the anti-hermitian part of the Fokker-Planck evolution can be written as a total derivative/pure gauge when $f^i$ is detailed balance.}

\paragraph{Supersymmetric Formulation.} Now we can see the supersymmetric formulation. Return to the path integral in \eqref{eq:stochPI}, and let us enforce the Langevin equation and Jacobian with a Lagrange multiplier and fermions i.e. write (up to prefactors):\footnote{In any stochastic problem, this is the place that a SUSY formulation can diverge. The stochastic problem generates $|\det\Delta|$ while the Faddeev-Popov determinant is signed $\det\Delta$. The two interpretations agree in cases when the sign of the determinant is fixed. For the Parisi-Wu/Langevin equation, our explicit determinant calculations show that the absolute value can be dropped (see Sections 4 \& 16 of \cite{zinn2021quantum} or \cite{Sethi:2026xhb}); formally, this follows from the fact that the Langevin SDE is a well-posed causal IVP and so has a unique solution $q(t)$ for each noise trajectory $\eta(t)$. 
By contrast, the Parisi-Sourlas/elliptic SPDE may admit multiple different solutions, so the solution must be studied in different regions of solution space separated by determinant zero modes, see also \cite{Kaviraj:2019tbg, Kaviraj:2020pwv, Rychkov:2023rgq}.}
\begin{align}
    \delta(E(q,\eta)) &= \int [\dd B]\, \exp(-\frac{1}{\lambda}\int_I \dd t\, B_i(t) E^i(t))\,,\\
    \det \Delta &= \int [\dd \psi \, \dd \bar{\psi}]\, \exp(\frac{1}{\lambda}\int \dd t\, \dd t'\, \bar{\psi}_i(t) \Delta^i_j(t,t') \psi^j(t'))\,.
\end{align}
The auxiliary bosonic integral over $B$ converges over the imaginary axis in complexified field space and the fermionic integrals are formal as always.\footnote{We choose $B$ to be imaginary to avoid carrying $i$ through derivations. In Examples \ref{sec:AModel} and \ref{sec:WZWModel} we will be much more cautious about removing/introducing $i$'s, especially in the \textit{boundary} actions.} 
Thus the stochastic path integral, with arbitrary insertions of $q$ in time, can be written
\begin{align}
    Z_{\mathrm{st}}[J]  
        &= \int [\dd\eta] [\dd q]\, \det\Delta \, \delta(E^i(q,\eta))\, e^{-\frac{1}{2\lambda}\int \dd t\, \eta^2 - \int \dd t Jq} \\
        &= \int [\dd\eta] [\dd q\, \dd \psi\, \dd \bar{\psi} \, \dd B]\, e^{-\frac{1}{2\lambda}\int \dd t\, \eta^2 - \int \dd t Jq - \frac{1}{\lambda}\int \dd t (B_i E^i - \bar{\psi}_i(\delta^i_j\partial_t + \frac{1}{2}\partial_j f^i)\psi^j)} \\
        &= \int [\dd q\, \dd \psi\, \dd \bar{\psi} \, \dd B]\, e^{- \int \dd t Jq - \frac{1}{\lambda}\int \dd t \left(B_i(\dot{q}^i + \frac{1}{2}f^i) - \frac{1}{2}B^2 - \bar{\psi}_i(\delta^i_j\partial_t + \frac{1}{2}\partial_j f^i)\psi^j\right)}\,.
\end{align}
For general $f^i$, the theory has an $\calN=1$ supersymmetry, and when $f^i = \partial^i S$, it has an $\calN=2$ supersymmetry (see Sections 16 and 17 of \cite{zinn2021quantum} or \cite{ovchinnikov2016introduction, Sethi:2026xhb} for more details in the general case). We call this \textit{Parisi-Sourlas-Wu supersymmetry} \cite{Nicolai:1979nr, Nicolai:1980jc, Parisi:1980ys, Cecotti:1981fu, parisi1982supersymmetric, cecotti1983stochastic}.\footnote{The conversion of random/stochastic systems to ``supersymmetric'' systems is quite general, see \cite{Kaviraj:2019tbg, Kaviraj:2020pwv, Kaviraj:2021qii, Rychkov:2023rgq, LeFloch:2025qjc} for recent works. However, supertranslations $Q$ do not always anti-commute to $H$ like they do here.} We will focus on a detailed balance force $f^i = \partial^i S$ going forward.

Our purported $\calN=2$ SUSY action is
\begin{equation}\label{eq:auxAction}
    S_{\mathrm{aux}} = \frac{1}{\lambda}\int_I \dd t \left(B_i(\dot{q}^i + \tfrac{1}{2}\partial^iS) - \tfrac{1}{2}B^2 - \bar{\psi}_i(\delta^i_j\partial_t + \tfrac{1}{2}\partial_j \partial^i S)\psi^j\right)\,.
\end{equation}
To see the supersymmetry, we rewrite this in terms of an $\calN=2$ superfield on the superspace $\calI :=I \times \bbR^{0|*2}$. See Appendix \ref{app:SupersymmetryFormulas} for our SUSY conventions. We denote a general superfield
\begin{equation}
    \Phi^i(t|\bar{\xi},\xi) 
        = q^i(t) - \xi \psi^i(t) + \bar{\xi}\bar{\psi}^i(t) + \xi \bar{\xi} B^i(t)\,,
\end{equation}
with superderivatives and differential representation of the supercharges given by:
\begin{equation}\label{eq:SUSYAlgebra}
    D 
        = \partial_{\xi} - 2\bar{\xi} \partial_{t} 
    \,,\quad
    \bar{D} 
        = \partial_{\bar\xi}
    \,,\quad        
    \calQ
        = \partial_{\xi}
    \,,\quad
    \calQb
        =\partial_{\bar\xi} + 2\xi \partial_t\,.
\end{equation}
The SUSY action $\delta \Phi := (\epsilon\!\calQ + \bar{\epsilon}\calQb)\Phi$ varies the superfield components by
\begin{equation}\label{eq:QQbarActions}
    \delta q^i 
        = - \epsilon \psi^i + \bar{\epsilon} \bar{\psi}^i\,,\quad
    \delta \psi^i
        = \bar{\epsilon}(2\dot{q}^i-B^i)\,,\quad
    \delta \bar{\psi}_i
        = - \epsilon B_i\,,\quad
    \delta B_i
        = 2 \bar{\epsilon} \dot{\bar{\psi}}_i\,.
\end{equation}
In superspace, the action $S_{\mathrm{aux}}$ becomes
\begin{equation}\label{eq:SSUSY}
    S_{\mathrm{aux}}[\Phi] = \frac{1}{2\lambda}\int_{\calI} [\dd t|\dd \bar{\xi} \dd\xi] \left(g_{ij}(\Phi)\bar{D}\Phi^i D\Phi^j + S(\Phi)\right)\,,
\end{equation}
where we more generally allow for the stochastic fields $q:I \to M$ to map into a general Riemannian manifold $(M,g)$. We note that the kinetic term and the potential term are separately SUSY invariant (up to boundary terms for $\bar{Q}$).

It is also useful to re-write this one more way, analogous to our unitarization $H_{\mathrm{FP},\pm} \mapsto H_{\pm}$. If we integrate out $B_i$ in $S_{\mathrm{aux}}$ using the equations of motion $B^i = \dot{q}^i + \partial^i S/2$, then the action becomes (for brevity, we only write the bosonic part and return to flat $M$)
\begin{align}
    S_{\mathrm{top}} 
        &:= \frac{1}{2\lambda} \int_I \dd t \, \left(\dot{q}^i + \tfrac{1}{2}\partial^i S\right)^2\\
        &= \frac{1}{2\lambda} \int_I \dd t \, \left(\dot{q}_i^2 + \tfrac{1}{4}(\partial_i S)^2\right) + \frac{1}{2\lambda}(S(t_f)-S(t_i))\\
        &=: S_{\mathrm{phys}} + \tfrac{1}{2\lambda}S(q)\big\vert_{q_i}^{q_f} \,.\label{eq:StopBd}
\end{align}
The name $S_{\mathrm{top}}$ comes from the fact that the action is actually topological after twisting. This can be seen by explicitly writing the object which trivializes the action. In the general off-shell form it is
\begin{align}
    S_{\mathrm{aux}} 
        &= \calQ V_{\mathrm{aux}}\,,\\
    V_{\mathrm{aux}} 
        &:= \frac{1}{\lambda}\int_I \dd t\, \bar{\psi}_i \left(-\dot{q}^i - \tfrac{1}{2}\partial^i S + \tfrac{1}{2}B^i \right)\,.
\end{align}

In summary, $S_{\mathrm{top}}$ is analogous to the Fokker-Planck frame, $S_{\mathrm{phys}}$ is analogous to the unitary frame, and $S_{\mathrm{aux}}$ is an auxiliary field form of $S_{\mathrm{top}}$.

\subsection{Generalities of \texorpdfstring{$\mathcal{N}=2$}{N=2} Supersymmetric Quantum Mechanics}\label{sec:SQMForBabies}
The action in \eqref{eq:SSUSY} describes an $\calN=2$ supersymmetric quantum mechanics with target manifold $(M,g)$. The original 0d potential $S$, defining the force $f^i = \partial^i S$, acts as a superpotential for this 1d system. Geometrically, the $q:I \to M$ are coordinates on $M$, and the fermions $\psi^i$ and $\bar{\psi}_i$ are sections of (the parity-shifted pull back) of $TM$ and $T^*M$ \cite{Witten:1982im} (see also \cite{Witten:2010zr, Gaiotto:2015aoa}). Famously, this gives a geometric interpretation to the space of states (as before, we proceed with notation below as if $M$ is flat, for simplicity).

Let us work with $S_{\mathrm{top}}$ with coordinates $(q^i, \psi^i)$. The conjugate momenta are\footnote{We factor out $\lambda$ from our momenta to have more conventional looking canonical commutators.}
$p_{i} = \dot{q}_i + \frac{1}{2}\partial_i S$ and $\pi_i = -\bar{\psi}_i$. To canonically quantize the theory, we demand
\begin{equation}
    [p_i, q^j] := - \lambda \delta_i^j
    \quad\text{and}\quad
    \{\pi_i,\psi^j\} = \lambda \delta_i^j\,.
\end{equation}
The SUSY charges become operators on the state space, given by
\begin{equation}
    Q_{\mathrm{top}} 
        = -\frac{1}{\lambda} \psi^i p_i 
        \,,\qquad
    \Qb_{\mathrm{top}} 
        = -\frac{1}{\lambda}\pi^i(p_i-\partial_i S) 
        \,.
\end{equation}
The commutation relations with fields are:
\begin{alignat}{5}
    [Q_{\mathrm{top}},q^i] 
        &= \psi^i \,,\quad
    &\{Q_{\mathrm{top}},\psi^i\} 
        &= 0 \,,\quad
    &\{Q_{\mathrm{top}},\bar{\psi}^i\} 
        &= \dot{q}^i + \tfrac{1}{2}\partial^i S\,,\\
    [\Qb_{\mathrm{top}},q^i] &= -\bar\psi^i \,,\quad
    &\{\Qb_{\mathrm{top}},\psi^i\} &= -\dot{q}^i + \tfrac{1}{2}\partial^i S \,,\quad
    &\{\Qb_{\mathrm{top}},\bar{\psi}^i\} &= 0\,.
\end{alignat}
This correctly matches minus the differential action in \eqref{eq:QQbarActions}. The Hamiltonian is:
\begin{align}
    H_{\mathrm{top}}
        &= -\frac{1}{2\lambda}\left(p^i(p_i-\partial_iS) - (\partial_i \partial_jS) \pi^j \psi^i\right)\\
        &= H_{\mathrm{FP}} + \frac{1}{4} \partial^2 S - \frac{1}{4\lambda} (\partial_i \partial_j S) [\bar{\psi}^j,\psi^i]\,.
\end{align}

A general formal wavefunction will be a function of the coordinates $(q^i,\psi^i)$. Identifying fermions on $M$ with differential forms, $\psi^i \sim \dd q^i$, gives an isomorphism
\begin{equation}
    C^\infty(\Pi TM) \cong \Omega^\bullet(M)\,.
\end{equation}
The momenta are therefore (locally) identified with $p_i \sim -\lambda \partial_i$ and $\bar{\psi}_i \sim - \lambda \iota_{\partial_i}$ (more generally, covariant derivatives on curved $M$). Thus, the formal space of states is $\Omega^\bullet(M)$, form degree is fermion number, and $Q_{\mathrm{top}}$ is the de Rham differential on $M$
\begin{equation}
    Q_{\mathrm{top}} \sim \dd q^i \partial_i\,.
\end{equation}

The ``physical theory'' corresponds to the action $S_{\mathrm{phys}}$, which differs from the topological theory by boundary terms, i.e., conjugating everything by $e^{S/2\lambda}$. For example, in the Hermitian/physical theory,
\begin{equation}
    p_i^{\mathrm{phys}} = e^{S/2\lambda} p_i e^{-S/2\lambda} = \lambda D_i^\dagger\,,
\end{equation}
as in \eqref{eq:covariantDerivative} (and up to our rescaling of momenta). The physical supercharges become the twisted de Rham differentials
\begin{align}
    Q_{\mathrm{phys}} 
        &= -\psi^i D_i^\dagger 
        \sim \dd - \frac{1}{2\lambda} \dd S \wedge\,,\\
    \Qb_{\mathrm{phys}} 
        &= -\bar{\psi}^i D_i
        \sim -\lambda\left(\dd^\dagger - \frac{1}{2\lambda} \iota_{\nabla S}\right)\,.
\end{align}
The physical Hamiltonian is
\begin{equation}
    H_{\mathrm{phys}} 
        = -\frac{1}{2}\{Q_{\mathrm{phys}}, \bar{Q}_{\mathrm{phys}}\} 
        = e^{S/2\lambda} H_{\mathrm{top}} e^{-S/2\lambda}\,.
\end{equation}
The physical Hilbert space $\calH_{\mathrm{SQM}}$ is identified with $L^2$-normalizable differential form valued wavefunctions on $M$
\begin{equation}
    \calH_{\mathrm{SQM}} \cong L^2\Omega^\bullet(M)\,.\label{eq:QMHilb}
\end{equation}

When SUSY is not broken, textbook arguments imply there exists an $E=0$ ground state, and that $E=0$ ground states are exactly the BPS states of the SQM -- annihilated by both supercharges $Q_{\mathrm{phys}}$ and $\Qb_{\mathrm{phys}}$. These are the harmonic representatives of the $Q_{\mathrm{phys}}$ cohomology classes in the state space $\calH_{\mathrm{SQM}}$. Consequently, we can write\footnote{Since the image of $\dd$ is not necessarily closed, we take the closure before computing cohomologies.}
\begin{equation}
    \calH_{\mathrm{BPS}} \cong H^\bullet(\calH_{\mathrm{SQM}}, Q_{\mathrm{phys}}) \cong H_{L^2}^\bullet(M, \dd_{-S/2\lambda})\,.
\end{equation}
On general smooth manifolds, the twisted and untwisted de Rham complexes are isomorphic, related by conjugation by $e^{S/2\lambda}$. The physical Hilbert space also has the $L^2$-normalizability condition, which can lead to different cohomology on non-compact manifolds depending on the difference of twisting functions $S_1 - S_2$ at infinity. We will elaborate on this in Section \ref{sec:NormalizabilityAnalyticCont}.

\subsection{SUSY Implications for the Langevin System}\label{sec:MorseTheory}
On an $n$-dimensional target manifold the SUSY Hamiltonians decompose over fermion number, e.g.
\begin{equation}
    H_{\mathrm{phys}} = \bigoplus_{k=0}^{n} H_{\mathrm{phys}}^{(k)}\,.
\end{equation}
Let $\ket{0}$ and $\ket{\Omega}$ denote the unfilled and filled fermion vacuum states, then, with our choice of canonical quantization conditions, $[\bar{\psi}_i, \psi^j]\ket{0} = -\lambda \delta_i^j \ket{0}$ and $[\bar{\psi}_i, \psi^j] \ket{\Omega} = + \lambda \delta_i^j \ket{\Omega}$, and so
\begin{equation}
    H_{\mathrm{phys}}^{(0)} = H_-\,,\quad
    H_{\mathrm{phys}}^{(n)} = H_+\,.\label{eq:ExtremeHamiltonians}
\end{equation}
Similar results hold for the topological/Fokker-Planck Hamiltonians.\footnote{We caution that in earlier literature, authors often consider only a single real bosonic variable $q:I \to \bbR$. Since the target is 1d, then $H_{\mathrm{phys}} = H_+ \oplus H_-$, we can relate retarded/advanced boundary conditions to periodic/anti-periodic boundary conditions for fermions on the circle $\det\Delta_{\pm} = \tfrac{1}{2}(\det\Delta_{\mathrm{AP}} \pm \det\Delta_{\mathrm{P}})$, and likewise relate the supersymmetric partition function to linear combinations of $Z_{\pm}$. This is obviously not true with more general targets and is refined by the SUSY analysis above (although, care should be taken to replace derivatives by covariant derivatives when appropriate).}

$H_{\mathrm{phys}}^{(0)}$ and $H_{\mathrm{phys}}^{(n)}$ have the formal zero-energy states $\Upsilon_{\mathrm{phys}}^{0}(q) = e^{S(q)/2\lambda}$ and $\Upsilon_{\mathrm{phys}}^{n}(q) = e^{-S(q)/2\lambda}$ respectively. As before, they are only formal because there is no guarantee that the ``state'' is actually normalizable when the target is non-compact. Embedded in the SUSY Hilbert space, these correspond to the formal SUSY ground states
\begin{equation}
    \ket*{\Upsilon_{\mathrm{phys}}^0} = 
    e^{+S(q)/2\lambda}\ket{0}\,,\quad
    \ket*{\Upsilon_{\mathrm{phys}}^n} = 
    e^{-S(q)/2\lambda}\ket{\Omega}\,,
\end{equation}
where the (formal) norms are:
\begin{align}
\norm*{\Upsilon^0_{\mathrm{phys}}(q)}^2 
    &= \int_M \dvol_M \, e^{+S(q)/\lambda} 
    =: \overline{Z}_{0d}\,,\\
    \norm*{\Upsilon^n_{\mathrm{phys}}(q)}^2 
    &= \int_M \dvol_M \, e^{-S(q)/\lambda} 
    =: Z_{0d}\label{eq:0dPF}\,.
\end{align}
On compact manifolds, both states are obviously normalizable for non-pathological $S$. On non-compact manifolds, this is not necessarily the case. One nice case is the case of a ``strongly confining'' potential, where $S \to \infty$ sufficiently quickly as $q \to \infty$ along all directions of $M$. By sufficiently quickly, we mean fast enough for our $L^2$-normalizability condition to give finite norms.

We can conjugate these states to the Fokker-Planck/topological perspective. For the $n$-fermion state, we have
\begin{align}
    \bra*{\Upsilon^n_{\mathrm{top}}} 
        &= \bra*{\Upsilon^n_{\mathrm{phys}}} e^{S(q)/2\lambda} 
        = 
        \bra*{\Omega}\,,\\
    \ket*{\Upsilon^n_{\mathrm{top}}} 
        &= e^{-S(q)/2\lambda} \ket*{\Upsilon^n_{\mathrm{phys}}} 
        = 
        e^{-S(q)/\lambda}\ket*{\Omega}\,.
\end{align}
This reproduces our results below \eqref{eq:eqmDist} after identification with forms: the right eigenvector of $H_{\mathrm{top}}^{(n)} = H_{\mathrm{FP}}$ with $0$ eigenvalue is the equilibrium distribution density $P_{\mathrm{eq}}(q,\psi) \propto e^{-S(q)/\lambda} \psi^1 \cdots \psi^n$, and the left eigenvector with $0$ eigenvalue is $\propto \bra{\Omega} \sim \star \dvol_M = 1$, i.e. the class of constant functions on $M$.

More generally, we have terms of all fermion number in $H_{\mathrm{phys}}$, whose ground/BPS states are in one-to-one correspondence with cohomology classes in $H^\bullet_{L^2}(M,\dd_{-S/2\lambda})$.\footnote{Naively, we anticipate a classical perturbative ground state for each nondegenerate critical point of the $S(q)$ potential, but these degeneracies are famously lifted by non-perturbative corrections \cite{Witten:1982im}.} A probabilistic interpretation of forms with degree $k < n$ is a bit tricky. One rough interpretation follows from the fact that $k$-forms can be integrated over lower dimensional $k$-chains $\calC \subset M$ in the configuration space, say locally defined by the $n-k$ constraints $q^{k+1} = \dots = q^n = 0$ in $M$. The integral of a $k$-form over $\calC$ could then roughly be interpreted as a conditional probability given the constraint. However, a major issue is that, unlike the top-form, we have no guarantee of positivity of the $k$-form over all of $\calC$ (see also \cite{Ovchinnikov:2012pb, Ovchinnikov:2012qa, ovchinnikov2016introduction, Ovchinnikov:2026jvd}), and so it does not define an actual probability distribution.

\paragraph{Morse Ground States and Boundary Conditions.} 
Let us temporarily restrict our attention to the topological/Fokker-Planck frame. We can prepare states at time $t=0$ by the standard Euclidean path integral. For example, to prepare the ground in-state $\ket*{\Upsilon^n_{\mathrm{top}}}$ at $t=0$, we simply perform the Euclidean path integral on $I = (-\infty,0]$ in the $n$-fermion sector with Hamiltonian $H_{\mathrm{top}}$. If we fix $q(0) = q_0$ and $\psi(0) = \psi_0$ at $t=0$, then the path integral prepares the wavefunction $\Upsilon^n_{\mathrm{top}}(q_0,\psi_0) = \braket*{q_0,\psi_0}{\Upsilon^n_{\mathrm{top}}}$. Later, we will consider the insertion of boundary observables.

In addition to this standard story, we can also consider interesting states from special Morse theory boundary conditions \cite{Witten:1982im, Witten:2010zr, Gaiotto:2015aoa}.  Since $S_{\mathrm{aux}}$ and/or $S_{\mathrm{top}}$ are $Q$-exact, we can easily solve them in the strict $\lambda \to 0$ limit. As $\lambda \to 0$, the action \eqref{eq:auxAction} clearly localizes onto solutions of the gradient flow equations
\begin{equation}\label{eq:gradFlow}
    \dot{q}^i + \frac{1}{2}\partial^i S = 0\,.
\end{equation}
We assume the superpotential $S$ has isolated non-degenerate critical points and suppose $p\in M$ denotes a classical critical point of $S$ (we leave more sophisticated $S$ to the references).

Now we place the theory on $I=[0,\infty)$. Let us consider solutions to the gradient flow equation \eqref{eq:gradFlow} which start at $q_0$ and end at $p$, i.e.
\begin{equation}\label{eq:WittenBC}
    q^i(t) = p^i \quad \text{as $t \to \infty$}\,.
\end{equation}
The (compactification of the) space of initial configurations $q_0$ ending at $p$ defines a chain $\calC_p \subset M$ with dimension $n-n_p$, where $n_p$ is the number of downward flowing eigenvalues of $\partial_i\partial_j S$ at $p$ (aka ``Morse index''). In the $\lambda \to 0$ limit, the path integral over fields satisfying \eqref{eq:WittenBC} localizes on a formal delta-function/distributional $n_p$-form out-state supported on $\calC_p$
\begin{equation}
    \braket*{\calC_p}{q_0,\psi_0} := \mathrm{PD}(\calC_p) \in \calD^{\prime,n_p}(M)\,,
\end{equation}
where $\mathrm{PD}(\calC_p)$ denotes the Poincar\'e dual of $\calC_p$ \cite{Witten:2010zr}. 

A priori, this state is not $Q$-closed and the boundary $\partial\!\calC_p$ can be decomposed over $\calC_q$ of Morse index $n_q = n_p + 1$. The exact differential is known to be
\begin{equation}
    \bra*{\calC_p}Q_{\mathrm{top}}
        = \sum_{q|n_q = n_p + 1} n_{qp} \bra*{\calC_q}
        \,,
\end{equation}
where the numbers $n_{qp}$ are the ``instanton numbers,'' counting oriented flows from $q$ to $p$ under downward gradient flow \cite{Witten:1982im, Gaiotto:2015aoa}. The SUSY interpretation is that $\calC_p$ is a classical perturbative ground state, but the differential $Q_{\mathrm{top}}$ knows about non-perturbative (i.e., global) structure of $M$ and modifies the BPS cohomology accordingly.

The left-eigenstate of $H_{\mathrm{FP}}$ with eigenvalue $0$ is equivalent in form-language to $1 \in \Omega^0(M)$. I.e., it is Poincar\'e dual to the manifold itself $\mathrm{PD}(M)$. Thus we will also denote it $\bra*{M}$.\footnote{For those taking score, this state is now called $\bra*{0_L}$, $\bra*{\Upsilon^n_{\mathrm{top}}}$, $\bra*{\Omega}$, and $\bra*{M}$ depending on context.} On compact $M$, or on non-compact $M$ with a confining $S$, we can decompose the chain $M$ into the stable basins formed by local minima $\calC_p$ with $n_p = 0$
\begin{equation}
    \bra*{M} = \sum_{p|n_p = 0} \bra*{\calC_p}\,.
\end{equation}
Of course, $\bra*{M} Q_{\mathrm{top}} = 0$, which reflects the fact that the signed boundary contributions of the $\calC_p$ cancel in the sum, despite not generically being $Q_{\mathrm{top}}$ closed themselves. This gives a decomposition of the 0d path integral, analogous to a thimble decomposition, in terms of Morse boundary conditions
\begin{equation}
    Z_{0d} = \braket*{M}{\Upsilon^n_{\mathrm{top}}} = \sum_{p|n_p = 0} \braket*{\calC_p}{\Upsilon^n_{\mathrm{top}}} \,.
\end{equation}
It would be interesting to understand how this structure changes in the presence of multiple equilibria, or if other $Q$-closed sums can appear.

\paragraph{Adding Boundary Observables.} We have discussed possibilities as $t\to\infty$, but we haven't commented on the behaviour of fields at $t=0$. Here we discuss the effect of turning on boundary local operators and boundary couplings in our path integral on $[0,\infty)$.

One choice of boundary condition is the Dirichlet boundary condition $q(0) = q_0$ and $\psi(0) = \psi_0$ at $t=0$. With Dirichlet boundary conditions, if we insert a $k$-form operator
\begin{equation}\label{eq:kFormOp}
    \calO(0) = \calO_{i_1\cdots i_k}(q(0)) \psi^{i_1}(0) \cdots \psi^{i_k}(0)\,,
\end{equation}
then it just evaluates to $\calO(q_0,\psi_0)$. E.g.,
\begin{align}
    \mel*{M}{\calO(0)}{q_0,\psi_0} 
        &= \int_{\substack{q(0)=q_0\\\psi(0)=\psi_0}} [\dd \Phi] \calO(q(0),\psi(0)) \, e^{-S_{\mathrm{top}}[\Phi]}\\
        &= \calO(q_0,\psi_0)\,.
\end{align}
More interesting Dirichlet boundary observables could be built using transverse (time) derivatives of the bulk fields at the boundary $\calO(p(0), \pi(0))$.

Alternatively, we can integrate over boundary values, and even turn on a boundary coupling $S_\partial$ at $t=0$. This ``enriched Neumann'' boundary condition corresponds to the in-state
\begin{equation}
    \ket*{S_\partial} = \int \dd q_0\, \dd\psi_0 \,e^{-S_\partial[q_0,\psi_0]/\lambda} \ket*{q_0,\psi_0}\,. 
\end{equation}
Note: the state $\ket{S_\partial}$ is generally \textit{not} $Q$-closed. At a Neumann boundary condition, the $k$-form operator \eqref{eq:kFormOp} will actually be an interesting observable. If we restrict the boundary coupling to only depend on $q_0$, i.e., $S_\partial(q_0,\psi_0) = S(q_0)$, a similar matrix element gives
\begin{align}
    \mel*{M}{\calO(0)}{S_\partial}
        &= \int [\dd \Phi] \calO(0) \, e^{-S_{\mathrm{top}}[\Phi] - \frac{1}{\lambda} S(q(0))}\\
        &= \int_M \calO(q_0,\psi_0) \, e^{-S(q_0)/\lambda}\label{eq:0dInt}\\
        &= Z_{0d} \, \expval*{\calO}_{0d}\,\quad\text{if $k = n$}.
\end{align}
Thus we see that the stochastic path integral prepares (the non-normalized) 0d correlation functions by using enriched Neumann boundary conditions on one side and the stochastic ground-state boundary condition on the other. Integrated on $M$, the only boundary insertions that evaluate to numbers/correlation functions are top-forms $\calO(0) \in \Omega^n(M)$. Lower degree forms $\calO(0)$ could potentially be integrated against out-states with higher-fermion number to obtain a non-trivial correlation function.

\subsection{BV Interpretation of the Boundary Data}\label{sec:BVNonsense}
There is a well-known BRST interpretation of the 1d bulk theory if we only care about the forward time-evolution $H_+$: we interpret the fields $q^i$ as elementary fields and interpret the superpartners $\psi^i$ and $\bar{\psi}_i$ as BRST ghosts (see e.g. \cite{Birmingham:1991ty, zinn2021quantum}). The $U(1)_R$ symmetry of the $\calN=2$ algebra can be viewed as $U(1)_\mathrm{gh}$ ghost number. Studying the $Q$-invariant $n$-fermion number SQM states $\Upsilon^n_{\mathrm{phys/top}}$ is studying states in ghost number $0$.

However, an even nicer interpretation is inherited by the 0d boundary observables and their correlation functions. Let us work in the topological picture and consider an enriched Neumann boundary condition $S_\partial$, with no $\psi_0$ dependence, and general Neumann boundary observable $\calO(0)$. Consider again the very general state:
\begin{equation}
    \ket{S_\partial; \calO} 
    := \calO(0) \ket{S_\partial}
    = \int \dd q_0 \,\dd\psi_0 \, e^{-S_\partial[q_0]/\lambda} \, \calO(q_0,\psi_0) \ket{q_0,\psi_0}\,.
\end{equation}
$Q_{\mathrm{top}}$ acts by $\dd$, which leads to the expression
\begin{equation}\label{eq:BPSState}
    Q_{\mathrm{top}} \ket{S_\partial; \calO} = \int \dd q_0\, \dd\psi_0\,  e^{-S_\partial[q_0]/\lambda}\, (\dd\!\calO - \tfrac{1}{\lambda}\dd S_\partial \wedge \calO) \ket{q_0,\psi_0}\,,
\end{equation}
To be $Q_{\mathrm{top}}$-closed, $\calO$ needs to be a twisted de Rham cocycle. We've already seen one easy way to satisfy this: have $\calO$ be a top-form. For top-forms, the localization of 1d amplitudes to 0d correlation functions is just usual supersymmetric localization of the amplitude between $Q$-closed states $\bra*{M}$ and $\ket*{S_\partial;\calO}$. 

In the 0d theory, $M$ plays the role of a ``space of fields,'' $q^i$ are coordinates on this space, and $\psi^i$ are (parity shifted) tangent coordinates. Wavefunctions of the 1d SQM are functions on $\Pi TM$, allowing us to effectively study the space of fields, forms, and integrals on it using (twisted) de Rham cohomology. For example, the twisted de Rham coboundary conditions
\begin{equation}\label{eq:DSEqn}
    \calO \sim \calO + (\dd\mathcal{X}-\tfrac{1}{\lambda}\dd\!\calS_{\partial}\wedge\mathcal{X})
\end{equation}
are just the Dyson-Schwinger equations for the $0$d theory.

Since we intend on moving beyond 0d/1d integrals to true infinite dimensional spaces of fields, we must recast this procedure. Even if we could define the infinite dimensional space $\calM$ of quantum fields, it would presumably be even harder to define $\lim_{k\to\infty} H^k(\calM)$ and compute integrals. Moreover, we don't normally think of observables in QFT as infinite dimensional forms on an infinite dimensional space.

The resolution to this problem is also suggested by the 1d SQM. A more natural frame is obtained by (fermionic) Fourier transforming to $\Pi T^*M$, i.e.
\begin{equation}
    \wt{\calO}(q,\chi) := \int \dd\psi \, e^{\chi_i \psi^i/\lambda} \calO(q,\psi)\,. 
\end{equation}
This is a change of polarization for the bulk SQM. In this frame, the bulk SUSY charge becomes
\begin{equation}
    \wt{Q}_{\mathrm{top}} = \lambda \frac{\partial}{\partial \chi_i} \frac{\partial}{\partial q^i}\,.
\end{equation}
Acting on the boundary state $\ket{S_\partial;\calO}$, the action of $Q_{\mathrm{top}}$ involves both the boundary couplings and local operator insertions (as above). This defines an effective operator on 0d boundary observables
\begin{equation}\label{eq:0dBV}
    \wt{Q}_{\mathrm{top}}^{0d} = \lambda \frac{\partial}{\partial q^i} \frac{\partial}{\partial \chi_i} - \frac{\partial S_\partial}{\partial q^i} \frac{\partial}{\partial \chi_i}  =: \Delta_{\BV}^{0d}\,.
\end{equation}
The operator $\wt{Q}_{\mathrm{top}}^{0d}$ is precisely the 0d quantum BV operator $\Delta_{\BV}^{0d}$. Applying the BV operator removes a $\chi_i$ from boundary observables and raises the BV ghost number by $+1$. The $\chi_i$ are the BV ghost number $-1$ anti-fields dual to the elementary fields $q^i$. Note that good 0d observables have BV ghost number $0$, which is the opposite of their 1d fermion number/form degree. In practice, dealing with top-forms on $M$ has been replaced by dealing with functions.

\paragraph{Langevin Processes as Absolutization of Relative QFTs.}
It is instructive to restate this point in the reverse: suppose we want to study correlation functions of a 0d theory with fields $q^i$ and classical action $S$. Our first job is to find the space of observables, i.e. all observables modulo redundancies from 0d equations of motion and symmetries. In the BV formalism, this is done by introducing anti-fields $\chi_i$ and writing the BV chain complex of all local operators built out of fields $q^i$ and anti-fields $\chi_i$. 

Naively, we would like to take something like $C^\infty(\Pi T^*M)$ as our space of all proto-observables. When $M$ is compact, oriented, and smooth, this is an okay choice, but when $M$ is non-compact, different choices of function spaces can give very different spaces of observables in cohomology. A somewhat universal choice is to consider observables built out of polynomials in the fields and anti-fields. Then the space of all formal observables is
\begin{equation}
    C_{\bullet}(M):= C^\infty_{\mathrm{poly}}(\Pi T^* M)\,,
\end{equation}
consisting of $k$-polyvector fields with polynomial coefficients in $q$ (polynomial is replaced by algebraic on general targets \cite{Kontsevich:2024esg}). This means $e^S$ itself is not an observable, for instance. 

With these subtleties in mind, the space of all formal observables is the ghost number $0$ cohomology of $\Delta_{\BV}^{0d}$ (with the same formula as \eqref{eq:0dBV}), i.e.
\begin{equation}\label{eq:0dObs}
    \mathrm{Obs}_{0d} = H^0(C_\bullet(M), \Delta_{\BV}^{\mathrm{0d}})\,.
\end{equation}
This is the space of all observables modulo Dyson-Schwinger equations for $S$. For example, if $M = \bbR$ and $S(q) = a q^4 + b q^3 + c q^2$, then the Dyson-Schwinger equations imply:
\begin{equation}
    4a \expval*{q^{n+3}} + 3b \expval*{q^{n+2}} + 2c \expval*{q^{n+1}} = \lambda n \expval*{q^{n-1}}\,,
\end{equation}
for all $n$ and any choice of state. Thus, all correlation functions can be recursively reduced to the normalization $\expval*{q^0}$ and moments $\expval*{q^1}$ and $\expval*{q^2}$.

In general, we see the data $q^i$ and $S$ do not actually specify the correlation functions of a QFT: only $S$-dependent consistency conditions between correlation functions. Instead of defining a partition function, or correlation functions, they only define a relative 0d QFT \cite{Freed:2012bs}, i.e. a boundary condition of a higher-dimensional QFT. In order to define an absolute QFT, we need some way to assign numbers to the 0d observables consistent with the relations between observables.

We saw a way to do this above: the 0d QFT is relative to a 1d bulk $\calN=2$ supersymmetric quantum mechanics $q: I \to M$ with superpotential $S$. The relative 0d theory defines a state $\ket*{S_\partial; \calO}$ of the SQM, and dual states $\bra{\Psi}$ in the quantum mechanics give different absolutizations of the relative QFT:
\begin{equation}\label{eq:absolutization}
    \expval*{\calO}_{S_\partial;\Psi} 
        := \frac{1}{Z_\Psi}\braket*{\Psi}{S_\partial; \calO}\,.
\end{equation}
The correlation function $\expval*{\calO_\alpha}_{S_\partial;\Psi}$ depends on the state $\Psi$, which in turn depends sensitively on bulk and boundary details of the path integral over $I$. However, since $\ket*{S_\partial;\calO}$ is $Q$-invariant when $\calO$ is a top-form (equivalently, a ghost-number $0$-observable in $\Delta_{BV}^{0d}$ cohomology), then if $\Psi$ is also $Q$-invariant, the overlaps become independent of bulk details by SUSY localization/$Q$-exactness of the action.

It is worth noting that we defined a relative theory to $\calN=2$ SQM, but we really only needed to define the theory as relative to the underlying 1d twisted/cohomological TFT. It is perhaps more accurate to say that a derived extension of the 0d QFT is relative to the full SQM. See \cite{BVThimTFT} for more details. This cohomological topological quantum mechanics formally admits a (cohomological) topological boundary state $\bra*{\Upsilon^n_{\mathrm{top}}} = \bra*{M}$. The Langevin procedure gives an interpretation of this state, and shows that the ``usual'' ``real'' path integral is obtained as the absolutization
\begin{equation}
    \expval*{\calO}_{0d,\mathrm{real}} := \frac{1}{Z_{0d}}\braket*{M}{S; \calO}\,.
\end{equation}

\subsection{Normalizability and Analytic Continuation}\label{sec:NormalizabilityAnalyticCont}
While our constructions of the previous sections were always formally valid, they depended on subtle convergence and normalizability properties. Central to this concern was the question of the convergence of the 0d partition function $Z_{0d}$ in \eqref{eq:0dPF}. Here we discuss some of the subtleties of non-compact manifolds. We continue to assume our manifold $M$ is oriented.

As a reminder, recall that singular cycles $C\in H_k(M)$ in ordinary homology are finite and pair with ordinary de Rham cohomology $\alpha \in H^k(M)$ to produce a number $\int_C \alpha$. Finiteness of cycles can be loosened to locally finite cycles $C^{\mathrm{BM}} \in H_{k}^{\mathrm{BM}}(M)$ in Borel-Moore homology, allowing for non-compact cycles as non-trivial homology elements. However, these potentially infinite chains can only be safely paired with compactly supported de Rham cohomology classes $\beta_c \in H_c^k(M)$ to produce a number $\int_{C^\mathrm{BM}} \beta_c$. Obviously, on compact manifolds, $H_k(M) \cong H^{\mathrm{BM}}_k(M)$ and $H^k(M) \cong H_c^k(M)$. Poincar\'e duality gives a perfect pairing between $H^k(M) \times H_c^{n-k}(M) \to \bbC$, and thus identifies $H^k(M) \cong H^{BM}_{n-k}(M)$ and $H_c^k(M) \cong H_{n-k}(M)$ \cite{Bott:1982xhp}.

When $M$ has a Riemannian metric, we can do away with some of these finiteness conditions by using the Hodge star. We equip the space of forms with an $L^2$-norm
\begin{equation}
    \braket*{\alpha}{\beta} = \int_M \beta\wedge\star\overline{\alpha}\,,
\end{equation}
which allows us to consider non-compactly supported -- but sufficiently rapidly decaying -- quantities at infinity, in exchange for some $g$-dependence. This is the $L^2$-norm defining our SQM Hilbert space in \eqref{eq:QMHilb}. On compact $M$, all smooth forms are $L^2$-normalizable, so the $L^2$-cohomology is isomorphic and simply changes the harmonic representatives \cite{Witten:1982im}. On a non-compact target, there is a priori no reason for the $L^2$-cohomology to match either $H^\bullet$ or $H^\bullet_c$, e.g., $H^0(\bbR^n) = \bbR$ generated by the constant class $[1]$, but this is not $L^2$-normalizable on $\bbR^n$ with $g=1$.

As noted, $\dd_{-S/2\lambda}$ is algebraically conjugate to $\dd$:
\begin{equation}
    \dd_{-S/2\lambda} = e^{S/2\lambda} \dd e^{-S/2\lambda}\,,
\end{equation}
but this is not generally an automorphism of $L^2\Omega^\bullet(M)$ since the $L^2$-domain changes. However, when we add the exponential strength twist $S$, then, for reasonable metrics, the asymptotic behaviour should be effectively determined entirely by the behaviour of $S$ at infinity. For sufficiently well-behaved $S$, the BPS state space is given by the relative cohomology \cite{dai2023witten}
\begin{equation}
    H_{L^2}^\bullet(M,\dd_{-S/2\lambda}) \cong H^\bullet(\overline{M}, \partial \overline{M})\,,
\end{equation}
where $\partial \overline{M}$ consists of regions in the compactification $\overline{M}$ upon which $S/\lambda \to \infty$.\footnote{We have intentionally avoided stating the precise mathematical definitions here because work on non-compact Morse theory (especially on real manifolds) is still a source of mathematical investigation. In the complex case we recommend \cite{dimca1993cohomology, pham1983vanishing, hien2008integral, johnson2015homological, fresan2018exponential, Kontsevich:2024esg} and in the real case \cite{braverman1997temperedcurrentscohomologyremote, farber2000witten} and especially Theorem 1.3 of \cite{dai2023witten} as starting points.} In practice, for two twisting functions $S_1$ and $S_2$ to give the same $L^2$-cohomology on a non-compact manifold $M$, it would be sufficient to show $S_1-S_2$ is bounded at infinity.

In particular, if $S/\lambda \to \infty$ as we approach $\partial\overline{M}$ from all directions (previously called ``confining''), then
\begin{equation}
    H_{L^2}^\bullet(M,\dd_{-S/2\lambda}) \cong H^\bullet_c(M)\,.
\end{equation}
E.g., on $M=\bbR$ with $S = x^2$, we have $\overline{M} = [-\infty,+\infty]$ and $\partial\overline{M} = \{-\infty,+\infty\}$, and there is a normalizable degree-$1$ ground state $\ket*{\Upsilon^1_{\mathrm{phys}}} = Z_{0d}^{-1/2}e^{-S/2\lambda} \ket*{\Omega}$ of $H_+$ -- this is the generator of $H_c^1(\bbR) = \bbR$. Note the importance of using compactly supported cohomology $H^1_c(\bbR)$, as $H^1(\bbR) = 0$. Conversely, for an anti-confining potential,
\begin{equation}
    H_{L^2}^\bullet(M,\dd_{-S/2\lambda})
    \cong
    H^\bullet(M)\,.
\end{equation}
Indeed, if $S=-x^2$, only the degree-$0$ state ground state $\ket*{\Upsilon^0_{\mathrm{phys}}} = \bar{Z}_{0d}^{-1/2}e^{S/2\lambda}\ket*{0}$ is normalizable, which generates the only non-zero class $H^0(\bbR) = \bbR$.

\paragraph{Analytic Continuation and the ThimTFT.} Generically, we do not even expect $M$ itself to be a class in the cohomology. I.e., we do not expect the integral $Z_{0d}$ to actually converge over $M$. One approach to studying convergence is to analytically continue the path integral: we complexify the space of fields, embedding $M \hookrightarrow M^{\bbC}$ and equipping $M^{\bbC}$ with a positive Hermitian (better, K\"ahler) metric, and analytically continue the action $S$ to a holomorphic function $S:M^{\bbC} \to \bbC$. A convergent analog of the partition function $Z_{0d}$ is any path integral
\begin{equation}\label{eq:compIntegral}
    Z_\Gamma := \int_{\Gamma\subset M^{\bbC}} {\dvol}^{\bbC}\, e^{-S(z)/\lambda}
\end{equation}
over a middle-dimensional cycle $\Gamma \in H^{\mathrm{rd}}_n(M^{\bbC},S(z)/\lambda)$ against a holomorphic top-form, locally $\dvol^{\bbC} = dz^1 \cdots dz^n$, which analytically continues the original real volume form \cite{pham1983vanishing, Frenkel:2006fy, Frenkel:2007ux, Frenkel:2008vz, Witten:2010cx, Witten:2010zr, johnson2015homological, liintroduction}. Elements of $H^{\mathrm{rd}}_n(M^{\bbC},S(z)/\lambda)$ are relative cycles in $M^{\bbC}$ which end in regions where $\Re(S(z)/\lambda) \to \infty$, up to homotopies which preserve the endpoint regions. A basis for the homology is given by the Lefschetz thimbles $\Gamma_p$, constructed in Section \ref{sec:MorseTheory}, attaching a thimble to the upwards flows of $\Re(S(z)/\lambda)$ through each critical point $p$. In the complexified case, all critical points have Morse index $n$, so each thimble state $\bra*{\Gamma_p}$ is automatically $Q$-closed. 

Since the fall-off is exponentially suppressed, any polynomial observable $\calO(z)$ (and presumably some meromorphic observables) can be integrated along $\Gamma \in H^{\mathrm{rd}}_n(M^{\bbC},S(z)/\lambda)$ to give a well-defined answer. These polynomial observables $\mathrm{Obs}_{0d,\mathrm{hol}}$ are the same ones which appear in the formal BV description \eqref{eq:0dObs}. Different rapid decay cycles $\Gamma$ provide different assignments of numbers $\expval{\cdots}_{\Gamma}$ to these observables, satisfying the same Dyson-Schwinger equations, i.e., for any $\Gamma$
\begin{equation}
    \frac{dS}{dz}\left[\frac{d}{dJ}\right]Z_\Gamma[J] - \lambda J Z_\Gamma[J] = 0\,.
\end{equation}
Again, the data of a correlation function is not specified by an observable $\calO(z)$ and an action $S(z)$ alone, but also a specific path integral contour $\Gamma$. As before, this can be interpreted in terms of relative QFTs: all path integral contours $\Gamma$ provide different absolute QFTs, and arise as different (topological) boundary conditions of the (topologically twisted) $\calN=2$ SQM with target $M^{\bbC}$. That is,
\begin{equation}
    Z_{\Gamma} = \braket*{\Gamma}{\Upsilon^n}\,,
\end{equation}
where $\ket*{\Upsilon^n} = e^{-S(z)/\lambda} \ket*{\Omega^{\bbC}}$ is the usual $S$-twisted in-state built on top of fully-filled \textit{holomorphic} fermions. We call this complexified theory that controls the space of thimbles the ``ThimTFT'' \cite{BVThimTFT}.

In the case of a complexified target $M^{\bbC}$, the integration of these polynomial (aka ``algebraic'') twisted de Rham $n$-forms $\omega$ against rapid decay $n$-cycles $\Gamma$ defines a perfect pairing of exponential periods \cite{kontsevich2001periods, hien2008integral, fresan2018exponential, Kontsevich:2024esg}:
\begin{equation}\label{eq:PeriodPairing}
    ([\omega], [\Gamma]) = \int_\Gamma \omega\, e^{-S(z)/\lambda}\,,
\end{equation}
generalizing Poincar\'e duality. Thus path integral contours $\Gamma$ are in one-to-one correspondence with linear functionals on formal BV observables and form the space of Dyson-Schwinger conformal blocks
\begin{equation}
    \mathrm{Obs}_{0d,\mathrm{hol}}^{\vee} \cong H_{n}^{\mathrm{rd}}(M^{\bbC},S(z)/\lambda; \bbC)\,.
\end{equation}
We leave the rest of this discussion to \cite{BVThimTFT}.\footnote{The Langevin equation has also been considered on complexified spaces in the context of Monte Carlo simulations and the sign problem \cite{Parisi:1983mgm, Klauder:1983zm} (see also \cite{Pehlevan:2007eq, Aarts:2009uq, Salcedo:2018fvt, Scherzer:2018hid, Scherzer:2019lrh, Hansen:2024kjm, Mandl:2024zvz, Boguslavski:2024yto, Mandl:2025mav, Asano:2025qfb, Mandl:2026vdc}). Above, we considered holomorphic observables $\calO(z)$ integrated against $e^{-S(z)/\lambda} \dvol^{\bbC}$ over middle-dimensional cycles $\Gamma$. Instead, the complex Langevin method considers holomorphic observables $\calO(z)$ integrated against a genuine probability density $P(z,\bar{z}) \geq 0$ on $M^{\bbC}$, coming from a Langevin equation on the $2n$-dimensional real manifold $M^{\bbC}$ with holomorphic drift
\begin{equation}\label{eq:ComplexLangevin}
    \dot{z}^i(t) = -\frac{1}{2}\partial_i S(z(t)) + \eta^i(t)\,,
\end{equation}
where $z^i = q^i + i r^i$ and real white-noise. The relevant question in the complex Langevin dynamics is then whether or not there exists a positive $P(z,\bar{z})$ on $M^{\bbC}$ that gives the correct answer on holomorphic observables.
The exact relation between the two methods is still an important open question \cite{Aarts:2013fpa, Nishimura:2017vav, Nishimura:2017eiu}, we discuss some potential connections in Example \ref{sec:AModel}.
}

\subsection{Comments on Fermions}\label{sec:FermionRant}
The SUSY perspective on the Langevin equation somewhat generalizes to fermions and fermionic SDEs. We again consider the action
\begin{equation}
    S_{\mathrm{aux}}[\Phi] = \frac{1}{2\lambda}\int_{\calI} [\dd t|\dd \bar{\xi} \dd\xi] \left(g_{AB}(\Phi)\bar{D}\Phi^A D\Phi^B + S(\Phi)\right)\,,
\end{equation}
but simply take our fields to have a target supermanifold $(M,g)$ with even and odd coordinates $X^A = (q^i, \theta^a)$. The superfields are now generally graded:
\begin{equation}
    \Phi^A(t|\bar\xi,\xi) := X^A(t) - \xi \Psi^A(t) + \bar{\xi} \bar\Psi^A(t) + \xi \bar\xi B^A(t)\,,
\end{equation}
with $\Psi^i$ fermionic, $\Psi^a$ bosonic, and so on. All algebraic expressions remain true by replacing every object by its appropriately graded counterpart. For example, $S_\mathrm{aux}$ is still $Q$-exact, with the same $V_{\mathrm{aux}}$ written in terms of the graded variables $X^A$. We highly recommend \cite{Witten:2012bg, Castellani:2017ycm} for introductions to supergeometry.

There are a number of technical problems with this approach, however. First up, we do not know how to make sense of $g_{ab}$ in the fermionic directions if there are not an even number of fermionic target directions. For example, we could consider the target supermanifold $\bbR^{1|1}$, but $g_{ab}$ will be degenerate in the fermionic direction. Next, the classical equation of motion will still be:
\begin{equation}
    \dot{X}^A + \frac{1}{2} g^{AB} \partial_B S(X) = 0\,,
\end{equation}
but this is hard to interpret for fermions. Specifically, the interpretation of fermionic solutions to this ODE, or spaces of solutions with some fixed boundary condition(s) analogous to the chains $\calC_p$, is far less clear. Said differently, there are no obvious thimbles for fermions, since fermionic integration is formal. 
Finally, if we perform a Nicolai map on our supermanifold-valued fields, we formally arrive at a theory of ``Grassmann odd Gaussian white noise'' with Grassmann odd Langevin equations:
\begin{equation}
    \dot{X}^a(t) = -\frac{1}{2}f^a(X(t)) + \eta^a(t)\,,
\end{equation}
subject to fermionic white noise
\begin{equation}
    \expval*{\eta^a(t)\eta^b(t')}_{\mathrm{st}} = \lambda g^{ab} \delta(t-t')\,,
\end{equation}
with higher moments defined by Wick contraction. In path integral language, this corresponds to the Grassmann noise integral
\begin{equation}
    \int [\dd\eta]\,  e^{-\frac{1}{2\lambda} \int \dd t\, \eta^a g_{ab} \eta^b}\,.
\end{equation}
Unlike the bosonic case, this does not technically define a good/positive probability measure, because the classical fermions are anti-commuting. Since they anti-commute, they do not fit into the standard framework of classical probability theory and must be formulated in the language of non-commutative probability theory.\footnote{Briefly, we can embed classical probability theory $(\Omega, P(x) \dd x)$ into a $C^*$-algebraic framework by replacing $\Omega$ with the commutative algebra of functions $\calA = L^\infty(\Omega)$ and the probability measure $P(x)\dd x$ by a positive linear functional $\omega$; in the usual way that classical probability theory embeds into quantum mechanics. Bosonic Gaussian random variables can be understood as $\calA_{\mathrm{bos}} = \mathrm{Sym}(\bbR^{n,\vee})$ with $\omega$ defining Wick's law for correlation functions. Fermionic Gaussian random variables are identical with $\calA_{\mathrm{fer}} = \extp^\bullet(\bbR^{n,\vee})$. Slightly ironically, our stochastic PDEs -- which are supposed to justify our construction of an interacting path integral measure -- require an independent non-commuting probability theory to deal with fermions.} An extremely rigorous approach to Grassmann random variables, associated finite and infinite dimensional Langevin equations, as well as a very thorough history of the subject in mathematics and physics, is presented in \cite{albeverio2022grassmannian}.

On the upside, the BV interpretation still survives. As we saw in Section \ref{sec:SQMForBabies}, there is a nice identification
\begin{equation}\label{eq:OmegaMSuper}
    \Omega^\bullet(M) \cong C_{\mathrm{poly.f}}^\infty(\Pi TM)\,,
\end{equation}
which holds on supermanifolds after restricting to functions with only polynomial dependence on the $\dd\theta^i$. Under this identification, $\dd q^i \sim \Psi^i$ and $\dd\theta^a \sim \Psi^a$ and so on. The polynomial restriction is necessary because $\dd\theta^a$ are even and can have arbitrarily high powers -- making top-forms ill-defined. Instead of integrating top-forms on a supermanifold $M$ of dimension $n|m$, we integrate ``integral top-forms,'' which locally have the form\footnote{The $\delta(\dd\theta^i)$ are Grassmann odd distributional forms. For example, when we write the usual Berezinian volume element $[\dd t|\dd\bar\xi \dd\xi]$ for our $\bbR^{1|2}$ worldline integrals, we are actually integrating a top integral form $\dd t \delta(\dd\bar\xi)\delta(\dd\xi)$. The Grassmann odd nature of the $\delta$-functions is compatible with the fact that the order of integration in Grassmann odd integrals matters \cite{Witten:2012bg, Castellani:2017ycm}.} 
\begin{equation}
    \calO = \calO(q,\theta) \dd q^1 \cdots \dd q^n \delta(\dd \theta^1) \cdots \delta(\dd\theta^m)\,.
\end{equation}
These integral top-forms $\calO$ are not contained in \eqref{eq:OmegaMSuper} -- they are distributions. Consequently, the polarization of the SQM where states appear as wavefunctions of $(X^A, \Psi^A)$ is not really natural once we are working on supermanifold targets. A similar issue occurs on infinite dimensional manifolds where even bosonic top-forms cannot be defined.

If we use a ``BV Polarization'' for the SQM Hilbert space, i.e. states are wavefunctions of $(X^A, \bm{\chi}_A)$ where $\bm{\chi}_A$ is Fourier dual to $\Psi^A \sim \dd X^A$. Under Fourier transform, the entire $\dd q^1 \cdots \dd q^n \delta(\dd\theta^1) \cdots \delta(\dd\theta^m)$ maps to $1$, thus the entire integral top-form just becomes the ghost number zero observable
\begin{equation}
    \wt{\calO}(X) \in C^\infty(\Pi T^*M)\,.
\end{equation}
Thus we still maintain our picture that the stochastic path integral is pairing a 0d BV cohomology class of ghost number $0$ with a reference out-state to define a 0d correlation function.

\section{Stochastic Quantization and Higher Dimensional Systems}\label{sec:higherDim}
In the previous section, we reviewed and refined the dictionary between stochastic (P)DEs and supersymmetric systems, generalizing known results and including Morse theoretic techniques. In this section, we turn to stochastic quantization and analyze it using the tools developed in the 0d/1d case, and discuss some illustrative examples.

In Section \ref{sec:StochasticGaussian} we review the stochastic quantization process of Parisi and Wu \cite{Parisi:1980ys} by coupling to pure Gaussian white noise and time evolving with the Langevin equation. By repeating identical manipulations to the 0d/1d case, we recast it as a supersymmetric process and discuss the localization argument for the equivalence to path integral quantization. We also introduce the Equilibrium TFT as the cohomological-topological field theory in which these amplitudes are computed. In Section \ref{sec:InterpolationProof} we give a Cardy-like interpolating action argument. In Example \ref{sec:AModel} we study the stochastic quantization of 1d quantum mechanics more carefully and, in particular, find a direct connection to analytic continuation and the A-model when formulated in a first-order form. In Example \ref{sec:WZWModel} we discuss the WZW model and the general procedure for attaching cohomological TFTs to collections of differential equations.

\subsection{Stochastic Quantization, Localization, and the EqmTFT}
\label{sec:StochasticGaussian}
As mentioned in the introduction, stochastic quantization proposes a scheme for quantizing a $d$-dimensional classical theory with action $S$ by coupling the $d$-dimensional theory to a heat bath and allowing the system to thermalize. The heat bath fluctuations are represented by some Markov stochastic process, and the principal claim is that these ($d+1$)-dimensional stochastic correlation functions become $d$-dimensional QFT correlation functions at late times. In the Parisi-Wu scheme, the heat bath is given by Gaussian white noise and time evolution by the Langevin equation. Consequently, the manipulations of this section will exactly mimic the results of Section \ref{sec:Stochastic}. Indeed, the main difference is now just that all previous expressions will carry spectator spatial directions, typically leading to supersymmetric non-relativistic theories with $z=2$ Lifshitz critical exponent (see e.g. \cite{Ardonne:2003wa, Dijkgraaf:2009gr, Chapman:2015wha, Boisvert:2025hex}).

Let us consider a $d$-dimensional Euclidean classical field theory of scalars $\phi^i:\bbR^d \to \bbR$ with classical action functional $S$. Say $i = 1,\dots, N$, and write $\Sigma = \bbR^d$ with target $M := \bbR^N$. The Parisi-Wu stochastic quantization of this $d$-dimensional theory is defined by the following procedure \cite{Parisi:1980ys, Damgaard:1987rr, zinn2021quantum}:
\begin{enumerate}
    \item Promote the classical $d$-dimensional fields to functions of the ``fictitious'' time $t$:
    \begin{equation}
        \phi^i(x) \mapsto \phi^i(x,t)\,.
    \end{equation}
    \item Couple the $\phi^i(x,t)$ degrees of freedom to white noise, with time-evolution defined by the detailed balance Langevin equation:
    \begin{equation}
        \pdv{\phi^i(x,t)}{t} = -\frac{1}{2}\!\!\left.\fdv{S}{\phi_i}\right\vert_{\phi(\cdot, t)} + \eta^i(x,t)\,.
    \end{equation}
    $\eta^i(x,t)$ is higher-dimensional Gaussian white noise, with two-point function
    \begin{equation}
        \expval*{\eta^i(x_1,t_1)\eta^j(x_2,t_2)}_{\mathrm{st}} = \lambda \delta^{ij} \delta^{(d)}(x_{12})\delta(t_{12})\,.
    \end{equation}
\end{enumerate}

\paragraph{Field Theory for the Langevin Process.} Our interest is in the stochastic correlation functions in the limit of large times. With general insertions of $\phi$ at equal times $\phi(\cdot, t) =: \phi_t$, we write the stochastic correlator:
\begin{equation}\label{eq:StochasticFPHigher}
    \stoch{F[\phi_t]} = \int [D\eta]\, F[\phi_t] e^{-\frac{1}{2\lambda} \int \dd t\, \dd^dx\, \eta^2(x,t)} = \int [D\phi_t]\, F[\phi_t] P[\phi_t,t;\phi_0, t_0]\,.
\end{equation}
As before, we view the conditional probability $P[\phi_t,t;\phi_0,t_0]$ as an overlap of states, and view the bulk ($d+1$)-dimensional theory as a non-relativistic non-Hermitian theory with Fokker-Planck Hamiltonian
\begin{equation}
    H_{\mathrm{FP}} = -\frac{\lambda}{2}\int_{\Sigma} \dd^dx\, \frac{\delta}{\delta \phi^i(x)}\left(\frac{\delta}{\delta \phi_i(x)} + \frac{1}{\lambda}\frac{\delta S}{\delta \phi_i(x)}\right)\,.
\end{equation}
This theory is generally PT-symmetric and, by conjugating all operators/states by $e^{S(\phi)/2\lambda}$, is equivalent to Hermitian Euclidean time-evolution generated by the Hamiltonian
\begin{equation}
    H_+ = \frac{\lambda}{2} \int_{\Sigma} \dd^dx\, \hat{D}_i^\dagger(x) \hat{D}_i(x)\,,\quad
    D^i(x) := \fdv{\phi^i(x)} + \frac{1}{2\lambda} \fdv{S}{\phi^i(x)}\,.
\end{equation}

A formal 0-energy ground state of $H_+$ is $\ket{0}$, given by the wavefunctional
\begin{equation}
     \Psi_0[\phi] = \braket{\phi}{0} = e^{-S[\phi]/2\lambda}\,.
\end{equation}
It is only formal because the state may not actually be $L^2$-normalizable. Let $\calM = \mathrm{Maps}(\Sigma, M)$ be the ``space of fields.'' Normalizability relies crucially on the finiteness of
\begin{equation}\label{eq:braket00}
    \braket{0}{0} = \int_{\calM} [D\phi]\, e^{-S[\phi]/\lambda} =: Z_{d}\,,
\end{equation}
which we recognize as the $d$-dimensional partition function.

As promised, this exactly mirrors the previous story, except with additional spectator spatial directions attached to everything. The general ($d+1$)-dimensional action is
\begin{equation}
    S_+ := \frac{1}{2\lambda} \int \dd t\, \dd^dx \left((\partial_t\phi)^2 - \frac{\lambda}{2}\frac{\delta^2 S}{\delta\phi_i^2} + \frac{1}{4}\left(\frac{\delta S}{\delta \phi^i}\right)^2\right)\,.
\end{equation}
If our starting $d$-dimensional theories are Lorentz/rotationally invariant, and the kinetic term of this theory is so that $\delta S/\delta \phi \sim (\partial^2)^m$, then the resulting ($d+1$)-dimensional theory will have one time-derivative and $2m$ spatial derivatives. Thus, the ($d+1$)-dimensional theory can also be described as an interacting deformation of a ($d+1$)-dimensional $z = 2m$ Lifshitz fixed point. 

\paragraph{Supersymmetric Formulation.} The stochastic PDE and ($d+1$)-dimensional field theory admit a supersymmetric formulation by identical manipulations to the 0d/1d case. However, since we do not have a Lorentz invariant theory, we are only guaranteed a $D=1$ $\calN=2$ SUSY algebra with spectator spatial directions. We note that we can consider this as a non-relativistic ($d+1$)-dimensional field theory \textit{or} as a $D=1$ $\calN=2$ SQM whose target is the infinite dimensional space of fields $\mathcal{M}$.\footnote{Warning, going forward we use $S(\Phi)$ for the action density (not integrated over $\Sigma$) interchangeably with the action itself. For example, in \eqref{eq:Auxd1} we mean the action density. Given that we will never mean to have two spatial integrations, no confusion should arise.}

The supersymmetric action is given in auxiliary field form by
\begin{equation}\label{eq:Auxd1}
    S_{\mathrm{aux}}[\Phi] = \frac{1}{2\lambda} \int_{\Sigma \times \calI} \!\!\!\! [\dd t \, \dd^dx | d\bar\xi d\xi] \left(\calG_{ij}(\Phi)\bar{D}\Phi^i D \Phi^j + S(\Phi)\right)
\end{equation}
with supercovariant derivatives as in \eqref{eq:SUSYAlgebra} and superfields promoted to depend on $x$ in the same way as before:
\begin{equation}
    \Phi^i(x,t|\bar\xi,\xi) := \phi^i(x,t) - \xi \psi^i(x,t) + \bar\xi \bar\psi^i(x,t) + \xi \bar\xi B^i(x,t)\,.
\end{equation}
Since all manipulations are identical to the 0d/1d case, just with additional spatial directions, we will suppress further details. We re-emphasize two points, as they will be important in Section \ref{sec:InterpolationProof}: both the kinetic term and potential term of \eqref{eq:Auxd1} are separately SUSY invariant (up to boundary terms), and both terms are $Q$-exact
\begin{align}
    S_{\mathrm{aux}} 
        &= \calQ V_{\mathrm{aux}}\,,\\
    V_{\mathrm{aux}} 
        &:= V_{\mathrm{kin}} + V_{\mathrm{pot}}\label{eq:Splitting}\\
        &= \frac{1}{\lambda}\int_{\Sigma \times I} \!\!\!\!\!\! \dd t\, \dd^dx\, \bar\psi_i\left(-\partial_t \phi^i + \frac{1}{2}B^i\right)
        - \frac{1}{2\lambda}\int_{\Sigma \times I} \!\!\!\!\!\! \dd t\, \dd^dx\, \bar\psi_i \frac{\delta S}{\delta\phi^i}\,.
\end{align}

The $\calN=2$ SUSY algebra gives an extremely formal interpretation to the space of states of the supersymmetric Lifshitz field theory as a supersymmetric quantum mechanics with configuration space $\calM$. In the $(\phi^i,\psi^i)$ polarization, the Hilbert space should consist of ``$L^2$-normalizable differential form-valued wavefunctions on $\calM$''
\begin{equation}
    \calH_{\mathrm{SQM}} \cong L^2 \Omega^\bullet(\calM)\,,\quad
    \calH_{\mathrm{BPS}} \cong H^\bullet_{L^2}(\calM, \dd_{-S/2\lambda})\,,
\end{equation}
where the physical SUSY operators are twisted de Rham operators
\begin{equation}
    Q_{\mathrm{phys}} = \int_{\Sigma} \dd^d x\, \psi^i \left(\frac{\delta}{\delta \phi^i} - \frac{1}{2\lambda}\frac{\delta S}{\delta \phi^i}\right)\,,\quad
    \Qb_{\mathrm{phys}} = -\int_{\Sigma} \dd^d x\, \bar\psi_i \left(\frac{\delta}{\delta \phi^i} + \frac{1}{2\lambda}\frac{\delta S}{\delta \phi^i}\right)\,.
\end{equation}
Likewise, the topological supercharge $Q_{\mathrm{top}}$ is the de Rham differential on $\calM$
\begin{equation}
    Q_{\mathrm{top}} = \int_{\Sigma} \dd^dx \,\psi^i\frac{\delta}{\delta \phi^i}\,.
\end{equation}

The Hamiltonians are the obvious generalizations of $H_{\mathrm{top}}$ and $H_{\mathrm{phys}}$ to the functional-integral case. Let us work in the topological/Fokker-Planck frame. If we fermionic Fourier transform $\psi^i \to \chi_i$ to the BV polarization, then our wavefunctionals become
\begin{equation}
    \widetilde{\Upsilon}[\phi,\chi] = \int [D\psi] \, e^{\frac{1}{\lambda}\int_{\Sigma} \dd^dx\, \chi_i \psi^i} \Upsilon[\phi,\psi]
\end{equation}
and in this polarization $Q_{\mathrm{top}}$ becomes
\begin{equation}
    Q_{\mathrm{top}} = \lambda \int_{\Sigma} \dd^dx\, \frac{\delta}{\delta \chi_i} \frac{\delta}{\delta \phi^i}\,.
\end{equation}
Formally, the ground state of the Fokker-Planck Hamiltonian is the ghost number $0$ state
\begin{equation}
    \ket*{\Upsilon_{\mathrm{top}}^\infty} = e^{-S[\phi]/\lambda} \ket*{0_{\mathrm{BV}}}\,,
\end{equation}
and the $0$-energy out-state is $\bra*{\Upsilon_{\mathrm{top}}^\infty} =:\! \bra*{\calM}$ formally satisfying $\braket*{\calM}{\Upsilon_{\mathrm{top}}^\infty} = Z_d$.

\paragraph{Stochastic Quantization, Localization, and the EqmTFT.} Stochastic quantization can now be phrased in terms of matrix elements and related to our previous formal machinery. If we take \eqref{eq:StochasticFPHigher} and general observable $\mathcal{O}$, then a general constant-time stochastic correlation function can be written
\begin{equation}
    \expval*{\calO[\phi_t]}_{\mathrm{st},\phi_0} 
    = \int [D\phi_t]\, \mel*{\phi_t}{\calO[\hat{\phi}] \,e^{-tH_{\mathrm{FP}}}}{\phi_0} 
    = \mel*{\calM}{\calO[\hat{\phi}] \, e^{-t H_{\mathrm{FP}}}}{\phi_0}\,.
\end{equation}
The main claim is that
\begin{equation}
    \expval*{\calO[\phi]}_{\mathrm{QFT}} := \lim_{t\to\infty} \expval{\calO[\phi_t]}_{\mathrm{st},\phi_0}
\end{equation}
behaves as a $d$-dimensional QFT correlation function which matches the usual $d$-dimensional path integral. But, if the system equilibrates at all, the right eigenstate of the Fokker-Planck Hamiltonian with 0-eigenvalue is just $\ket{\mathrm{eq}} = \ket*{\Upsilon_{\mathrm{top}}^\infty}$, thus at $t=0$ on the semi-infinite line $[0,\infty)$ we replace $e^{-tH_{\mathrm{FP}}}\ket{\phi_0}$ by $\ket*{\Upsilon^{\infty}_{\mathrm{top}}}$. In other words, we assume our in-state to be
\begin{equation}
    \ket*{S;\calO_1 \cdots \calO_n} = e^{-S[\hat{\phi}]/\lambda} \calO_1(x_1,0) \cdots \calO_n(x_n,0)\ket{0_{\mathrm{BV}}}\,,
\end{equation}
where $\calO_i(x_i,0)$ are ghost number $0$ observables at the boundary, i.e. built only out of $\phi$. Now we take the overlap with $\bra*{\calM}$:
\begin{align}
    \braket*{\calM}{S;\calO_1 \cdots \calO_n}
        &= \int [D\Phi] \calO_1(x_1,0)\cdots \calO_n(x_n,0) e^{-S_{\mathrm{top}}[\Phi]-\frac{1}{\lambda}S[\phi(0)]}\\
        &= \int_{\calM} [D\phi]  \calO_1(x_1)\cdots \calO_n(x_n) e^{-S[\phi]/\lambda}\\
        &= Z_d \expval{\calO_1(x_1) \cdots \calO_n(x_n)}
\end{align}
This is the main claim of stochastic quantization: \textit{assuming equilibration, the overlap of bulk $(d+1)$-dimensional stochastic states produces standard $d$-dimensional path integral quantized correlation functions}.

As before, we note the important role played by the enriched Neumann boundary state in encoding the dynamics of the theory via the Schwinger Dyson equations on the otherwise $S$-independent algebra of observables. Consider a general polynomial boundary observable built out of fields and anti-fields $\calO[\phi,\chi]$, then
\begin{equation}
    Q_{\mathrm{top}}\left(e^{-S[\phi]/\lambda} \calO[\phi,\chi]\right)
        = e^{-S[\phi]/\lambda} Q_S \calO[\phi,\chi]\,,
\end{equation}
where
\begin{equation}
    Q_S = \int_{\Sigma} \dd^dx\left(\lambda \frac{\delta}{\delta\phi^i} - \frac{\delta S}{\delta\phi^i}\right)\frac{\delta}{\delta\chi_i}\,.
\end{equation}
Thus, the condition that the state $\ket{S;\calO}$ is $Q_{\mathrm{top}}$-closed demands that boundary operators satisfy
\begin{equation}
    Q_S \!\calO = 0\,,
\end{equation}
with two observables identified if $\calO \sim \calO + Q_S \mathcal{X}$. Thus $Q_{S}$ is a $d$-dimensional BV operator.\footnote{Now that we have added spatial directions, there are of course UV divergences from coincident operators. They can be treated similarly to any UV divergences in QFTs, as described in detail in \cite{zinn2021quantum}.}

Since we only study the overlap of $Q_{\mathrm{top}}$-closed states, and the bulk action is $Q$-exact, the amplitude is independent of the ($d+1$)-dimensional bulk path integral, and so prepares the usual $d$-dimensional path integral without any dependence on the length of the original interval. This is the localization argument for the equivalence of standard path integral quantization to stochastic quantization.\footnote{One might complain that the left and right ground states of the Fokker-Planck Hamiltonian were only formal, with a potentially infinite norm $Z_d$. If this is the case, then even the usual $d$-dimensional path integral is not finite. This is another motivation to consider analytic continuation in more depth. More optimistically, we could try to understand the convergence of $Z_d$ by understanding the space of states of the SQM on $\calM$, which in turn reduces to understanding the cohomology of the infinite dimensional space of fields. At least in some cases (e.g. the space of 2d flat connections) this may be possible. See also Example \ref{sec:WZWModel}.}

If we focus only on the cohomological/topologically twisted $\calN=2$ SQM, and place the theory on an interval $[0,1]$ with $\bra*{\calM}$ prepared at $t=1$ as a topological boundary condition, then stochastic quantization describes a sandwich construction or SymTFT \cite{Kong_2017, Freed:2018cec, Gaiotto:2020iye, Apruzzi:2021nmk}, where the out-state $\bra*{\calM}$ corresponds to a reference boundary condition, and the original action and $d$-dimensional observables (viewed as boundary observables) correspond to an enriched Neumann state.\footnote{We do not use the words physical and topological boundary condition because physical and topological were already used to refer to forms of the SQM action. Sorry!} We call this bulk cohomological theory the Equilibrium TFT (``EqmTFT''), see Figure \ref{fig:symtft_wavefunction_overlap}.

\begin{figure}[t]
    \centering
    \begin{tikzpicture}[thick]
    
    \def\Depth{-3.2}
    \def\Height{2.7}
    \def\Width{2.7}
    \def\Future{-1.5}
    
    \begin{scope}[local bounding box=twodpic]
    
    \coordinate (O) at (0,0,0);
    \coordinate (A) at (0,\Width,0);
    \coordinate (B) at (0,\Width,\Height);
    \coordinate (C) at (0,0,\Height);
    
    \draw[black, fill=purple!18, opacity=0.85] (O) -- (A) -- (B) -- (C) -- cycle;
    \draw[black] (O) -- (A) -- (B) -- (C) -- cycle;
    
    \coordinate (p1) at (0,0.25*\Width,0.40*\Height);
    \coordinate (pn) at (0,0.76*\Width,0.48*\Height);
    \node[
        circle,
        fill,
        inner sep=1.5pt,
        label={[xshift=-8pt,yshift=-12pt]:$\mathcal O_n$}
    ] at (p1) {};
    \node[scale=1] at (0,0.525*\Width,0.45*\Height) {$\vdots$};
    \node[
        circle,
        fill,
        inner sep=1.5pt,
        label={[xshift=-8pt,yshift=-12pt]:$\mathcal O_1$}
    ] at (pn) {};
    \end{scope}
    
    \node[anchor=north, yshift=-8pt] at (twodpic.south)
        {$\expval{\mathcal O_1 \cdots \mathcal O_n}_{\textcolor{red!45!purple!85!black}{\calM}}$};
    
    \begin{scope}[xshift=9.7cm]
    
    \coordinate (O) at (0,0,0);
    \coordinate (A) at (0,\Width,0);
    \coordinate (B) at (0,\Width,\Height);
    \coordinate (C) at (0,0,\Height);
    
    \coordinate (D) at (\Depth,0,0);
    \coordinate (E) at (\Depth,\Width,0);
    \coordinate (F) at (\Depth,\Width,\Height);
    \coordinate (G) at (\Depth,0,\Height);
    
    \coordinate (Dinf) at (\Depth+\Future,0,0);
    \coordinate (Einf) at (\Depth+\Future,\Width,0);
    \coordinate (Finf) at (\Depth+\Future,\Width,\Height);
    \coordinate (Ginf) at (\Depth+\Future,0,\Height);
    
    \begin{scope}[local bounding box=bulkpic]
    
    \fill[white] (O) -- (A) -- (Einf) -- (Dinf) -- cycle;
    \fill[white] (C) -- (B) -- (Finf) -- (Ginf) -- cycle;
    \fill[white] (O) -- (C) -- (Ginf) -- (Dinf) -- cycle;
    \fill[white] (A) -- (B) -- (Finf) -- (Einf) -- cycle;
    
\pgfdeclarehorizontalshading{bulkfade}{100bp}{
    color(0bp)=(purple!28);
    color(25bp)=(purple!28);
    color(29bp)=(purple!22);
    color(34bp)=(purple!12);
    color(40bp)=(purple!4);
    color(46bp)=(white);
    color(75bp)=(white);
    color(100bp)=(white)
}

\shade[shading=bulkfade]
    (O) -- (A) -- (Einf) -- (Dinf) -- cycle;
\shade[shading=bulkfade]
    (C) -- (B) -- (Finf) -- (Ginf) -- cycle;
\shade[shading=bulkfade]
    (O) -- (C) -- (Ginf) -- (Dinf) -- cycle;
\shade[shading=bulkfade]
    (A) -- (B) -- (Finf) -- (Einf) -- cycle;
    
    \begin{scope}[local bounding box=finitebdry]
    \draw[black, fill=white] (O) -- (A) -- (B) -- (C) -- cycle;
    
    \coordinate (p1) at (0,0.25*\Width,0.40*\Height);
    \coordinate (pn) at (0,0.76*\Width,0.48*\Height);
    \node[
        circle,
        fill,
        inner sep=1.5pt,
        label={[xshift=-8pt,yshift=-12pt]:$\mathcal O_n$}
    ] at (p1) {};
    \node[scale=1] at (0,0.525*\Width,0.45*\Height) {$\vdots$};
    \node[
        circle,
        fill,
        inner sep=1.5pt,
        label={[xshift=-8pt,yshift=-12pt]:$\mathcal O_1$}
    ] at (pn) {};
    \end{scope}
    
    \draw[gray] (O) -- (D);
    \draw[black] (A) -- (E);
    \draw[black] (B) -- (F);
    \draw[black] (C) -- (G);
    
    \draw[dashed,gray] (D) -- (Dinf);
    \draw[dashed] (E) -- (Einf);
    \draw[dashed] (F) -- (Finf);
    \draw[dashed] (G) -- (Ginf);
    
    \node at (\Depth/2+\Future/2,\Width/2,\Height/2) {$\mathcal N=2$ SQM};
    
    \end{scope}
    
    \node[anchor=north, yshift=-8pt] at (finitebdry.south)
        {$\ket{S; \mathcal O_1\cdots\mathcal O_n}$};
    
    \begin{scope}[local bounding box=inftybdry]
        \path (Dinf) -- (Einf) -- (Finf) -- (Ginf) -- cycle;
    \end{scope}
    
    \node[anchor=north, yshift=-8pt, xshift = -8pt] at (inftybdry.south)
        {$\bra{\textcolor{red!45!purple!85!black}{\calM}}$};
    
    \end{scope}
    
    \node at ($(twodpic.east)!0.5!(bulkpic.west)$) {$\Longleftrightarrow$};
    
    \end{tikzpicture}
    \caption{Left, the two-dimensional path integral over $\calM$ prepares the correlation function $\expval*{\calO_1\cdots\calO_n}_{\calM}$. Right, this correlation function can be equivalently represented as an overlap of states $\bra*{\calM}$ and $\ket*{S;\calO_1\cdots\calO_n}$ of a $(2+1)$-dimensional bulk $\calN=2$ SQM or a cohomological-topological EqmTFT. The enriched Neumann boundary state $\ket*{S;\calO_1\cdots,\calO_n}$ carries boundary insertions matching the $2$-dimensional theory.}
    \label{fig:symtft_wavefunction_overlap}
\end{figure}

\subsection{Cardy's Interpolation Argument}\label{sec:InterpolationProof}
An interpolation argument for the equivalence of stochastic quantization to usual path integral quantization was developed by Cardy in the context of Parisi-Sourlas codimension-2 reduction in \cite{Cardy:1983aa} (see also \cite{LeFloch:2025qjc}). One can try to adapt the arguments to the codimension-1 stochastic quantization case (see e.g. \cite{Kirschner:1986zh, Gozzi:1984au, Damgaard:1987rr}). To the best of our knowledge, all of these proofs explicitly or implicitly rely on equilibration as $t\to\infty$ (or callous treatment of boundary terms), in which case we may as well replace the state $e^{-T H_{\mathrm{FP}}}\ket{\phi_0}$ with $\ket{\Upsilon}$ again, then the previous formulations make it obvious that it formally recovers path-integral quantization. Here we will discuss issues of boundaries of supermanifolds more carefully and present the localization argument in a path-integral language as a demonstration.\footnote{It would be very interesting to us to produce a Cardy-like argument for interpolation, that would presumably rely on interpolating between the Dirichlet boundary condition $\ket{\phi_0}$ and the enriched Neumann boundary condition.}

\paragraph{Some Basic Supergeometry.} Thus far, we have been somewhat heuristic about our supermanifold with boundary $\Sigma \times \calI$, with $t\in[0,\infty)$. One way to define our supermanifold with boundary, heuristically called $\Sigma \times \calI$, is by embedding it into $\calY := \Sigma \times \bbR^{1|*2}$. Different choices of embedding can lead to the same reduced (bosonic) manifold, but define different supermanifolds. Let $f(x,t|\bar\xi,\xi)$ be a globally well-defined function on $\calY$ with the property that $f(x,t|0,0)$ is real, and modulo rescalings of the form $f \mapsto e^{\varphi} f$ where $\varphi(x,t|0,0)$ is real. We can define a supermanifold with boundary by the condition $f \leq 0$; in practice, $\calM_f$ is defined so that any compactly supported integral top-form $\sigma$ on $\calY$ satisfies
\begin{equation}
    \int_{\calM_f} \sigma = \int_{\calY} \Theta(-f)\sigma\,.
\end{equation}
With this definition, every
\begin{equation}
    f_b(x,t|\bar\xi,\xi) := -t + b \bar\xi\xi
\end{equation}
defines a different supermanifold with boundary $\calM_b$. If we consider the most general integral top-form (with compact support as $t \to \infty$) 
\begin{equation}
    \sigma = [\dd t \dd^d x|\dd \bar\xi \dd\xi](\sigma_{00}+\xi \sigma_{10} + \bar\xi \sigma_{01} + \xi\bar\xi \sigma_{11})\,,
\end{equation}
then
\begin{equation}
    \int_{\calM_b} \sigma = \int_{\calY} \Theta(-f_b)\sigma = \int_{\Sigma \times I} \dd t \dd^dx \, \sigma_{11}(x,t) + b \int_{\Sigma} \dd^dx \, \sigma_{00}(x,0)\,.
\end{equation}
Thus, subtly different supermanifolds give different boundary terms when integrating forms, even though the reduced manifolds are identical!

The $\calN=2$ $D=1$ superalgebra is not the isometry algebra of $\bbR^{1|*2}$ with the naive flat metric, it is the isometry algebra of $\bbR^{1|*2}$ with some off block-diagonal/super-torsion elements \cite{Castellani:2017ycm, BVThimTFT}. On our half-line $\calM_b$, we expect some of these isometries to be broken. Using the differential representations of the generators \eqref{eq:SUSYAlgebra}, a generator is preserved if it maps $f_b$ to itself, up to the aforementioned rescaling identifications. Of course, time translations $\calH$ are never preserved, and $\calQ$ is preserved iff $b=0$, and $\calQb$ is preserved iff $b=2$. In hindsight, this is not surprising: in Appendix \ref{app:SupersymmetryFormulas} we already judiciously chose to work in chiral coordinates where the $\calQ$-preserving supermanifold was naturally given by $f_0(t) = -t$. 

In the following, we work on the finite strip with reduced manifold $\Sigma \times [0,T]$ defined by
\begin{equation}
    \calM_T = \{f_0 \leq 0, f_T \leq 0 \,|\, f_0 = -t\,, f_T = t-T\}\,.
\end{equation}
For integral forms, this means
\begin{equation}
    \int_{\calM_T} \sigma = \int_{\calY} \Theta(t)\Theta(T-t) \sigma\,,
\end{equation}
and we will eventually consider $T \to \infty$. $\calM_T$ clearly preserves $\calQ$.

\paragraph{An Interpolating Action.} Working on $\calM_T$ we write down an interpolating action:
\begin{align}
\begin{split}
        S_{s}[\Phi] 
        := S_{\mathrm{kin}}[\Phi] + s S_{\mathrm{pot}}[\Phi]
        &+ \frac{1}{\lambda}\left[S[\phi(x,0)] - \int_{\Sigma} \dd^dx\, \bar\psi^i \psi^i\Big\vert_{t=0}\right]
\end{split}
\end{align}
where $s \in [0,1]$. Since the bulk kinetic and potential term individually preserve supersymmetry, there is no bulk breaking of SUSY from deforming the parameter $s$. Thus, to preserve supersymmetry, we must only track the effect of the boundary terms. The explicit boundary variation of the interpolating actions is:
\begin{align}
    \delta S_{s}[\Phi]\Big\vert_{t=0}
        &= \frac{1}{\lambda}\int_{\Sigma}\dd^dx \left[\left(-B_i + \tfrac{\delta S}{\delta\phi^i}\right)\delta\phi^i - \delta\bar{\psi}_i\psi^i\right]\,,\\
    \delta S_{s}[\Phi]\Big\vert_{t=T}
        &= \frac{1}{\lambda}\int_{\Sigma}\dd^dx \left[B_i \delta\phi^i - \bar{\psi}_i\delta\psi^i\right]\,.
\end{align}

Our setup should have free $\phi(x,0)$ and $\bar{\psi}(x, 0)$ boundary fields at $t=0$, i.e., Neumann boundary conditions for $\phi$ and the $\bar{\psi}$-polarization at $t=0$. Thus we get
\begin{equation}
    B_i(x,0) = \frac{\delta S}{\delta\phi^i}\Big\vert_{t=0}\,,\qquad
    \psi^i(x,0) = 0\,.
\end{equation}
This is precisely the $Q$-supersymmetric ``enriched BV Neumann'' boundary condition prepared in the previous section. Likewise, the $t=T$ version of the reference state $\bra{\calM}$ should have free $\phi$ but zero $\bar{\psi}$-number. Thus we have
\begin{equation}
    B_i(x,T) = 0\,,\qquad
    \bar{\psi}^i(x,T) = 0\,.
\end{equation}
These boundary conditions are also $Q$-supersymmetric. Using these chiral supersymmetric boundary conditions on the strip $\calM_T$, the action is fully $Q$-supersymmetric for all $s$ -- interpolating between a free bulk and interacting bulk, with one enriched Neumann boundary in the BV-polarization and one Neumann boundary condition in the de Rham polarization.


Now we write the path integral coupled to a source $\calJ_i(x,t|\bar\xi,\xi):= \delta_\partial^{(1|2)}(t|\bar\xi,\xi) J_i(x)$ for boundary degrees of freedom:
\begin{equation}
    Z_{s}[\calJ;\calM_T] := \int [D\Phi] \exp(-S_{s}[\Phi] + \int_{\calM_T} [\dd t \dd^dx|\dd\bar\xi\dd\xi] \calJ_i(x,t|\bar\xi,\xi) \Phi(x,t|\bar\xi,\xi))\,.
\end{equation}
With our boundary condition, this source-term $\int\calJ\Phi$ is automatically $Q$-invariant. At $s=0$ the bulk action is free, and we can integrate out the bulk fields, leaving only the $d$-dimensional path integral up to a $J$ and $\phi$-independent normalization constant:
\begin{align}
    Z_0[\calJ;\calM_T] 
        &= C_T \int [D\phi(x,0)] \exp\left(-\frac{1}{\lambda}S[\phi(x,0)] + \int_{\Sigma} \dd^dx J_i(x) \phi^i(x,0)\right) \\
        &= C_T Z_{\mathrm{QFT}}[J; \Sigma]\,.
\end{align}
At $s=1$, we have the path integral for our stochastic quantization with insertions. The Cardy action interpolates between the two. Now we take the generating function for normalized correlation functions and consider its derivative as a function of $s$
\begin{align}
\begin{split}
    \partial_s \log \frac{Z_s[\calJ;\calM_T]}{Z_s[0,\calM_T]}
        &=  - \expval*{\calQ \!V_{\mathrm{pot}}}_{s,J} 
            +  \expval*{\calQ \!V_{\mathrm{pot}}}_{s,0}
        = 0\,,
\end{split}
\end{align}
by supersymmetry of the action, boundary conditions, and source terms.

\subsection{Example: Quantum Mechanics and the A-Model}\label{sec:AModel}
The previous sections considered stochastic quantization as SQM on an infinite dimensional manifold $\calM$ in a somewhat formal way. It is instructive to work through the process with a concrete example in greater than 0d and in a first-order formalism.

\paragraph{Second Order Form.} Let us consider a 1d theory of scalars $q:\Sigma \to M$. We will take $\Sigma = S^1$ and let $M$ be the configuration space with Riemannian metric $g_{ij}(x)$. Then the space of field configurations is the loop space $\calM = \mathrm{Maps}(\Sigma,M) = LM$ and inherits a Riemannian metric $\calG_{(i,x)(j,x')} = g_{ij}(q(x)) \delta(x-x')$ from $M$. We take $g_{ij} = \delta_{ij}$ and our 1d action as
\begin{equation}
    S[q] = \int \dd x\, \left(\frac{1}{2} (\partial_xq)^2 + V(q)\right)\,.
\end{equation}
The Langevin equation says that a classical loop evolves towards some homotopic configurations by gradient flow, with white noise on top:
\begin{equation}
    \partial_t q^i(x,t) = \frac{1}{2}\left[\partial_x^2 q^i(x,t) - \partial_i V(q(x,t))\right] + \eta^i(x,t)\,.
\end{equation}

If we suppress $x$ (or think of it as a label for fields), then the correspondence between stochastic PDEs and SQM says that this is equivalent to an $\calN=2$ SQM with target $\calM = LM$, or, if we expand out the loop parameter direction, then it is a $2d$ sigma model with target $M$ and $D=1$ $\calN=2$-type supersymmetry in time:
\begin{equation}\label{eq:topQM}
    S_{\mathrm{top}}
        = \frac{1}{\lambda}\int \dd t \, \dd x\, \left[\frac{1}{2}\left(\partial_t q^i - \frac{1}{2}\partial_x^2 q^i + \frac{1}{2}\partial_i V\right)^2 - \bar{\psi}_i \left(\delta^i_j \partial_t - \frac{1}{2}\delta^i_j \partial_x^2 + \frac{1}{2}\partial_i\partial_j V\right)\psi_j\right]
\end{equation}
Since the original theory was second order in Euclidean time $x$, the resulting 2d sigma model is an interacting deformation of a free $z=2$ Lifshitz theory of bosons and fermions. 

If at this point we took $S[q]$ to be quadratic/harmonic oscillator, then the Langevin equation could be solved identically to \eqref{sec:BrownianExample} for fixed $x$ and $\eta$ (or better, work in Fourier space in $x$). Then the stochastic correlation functions recover finite-temperature correlation functions in the QM.

\paragraph{First Order Form.} We can consider the same classical theory, but presented in a first-order form, i.e. with two sets of scalars $X^A = (q^i,p_j):\Sigma \to T^*M$. In principle, this should give us a $z=1$ relativistic bulk presentation of the same theory. The configuration space is now $T^*M$, which has a Riemannian metric $G$ inherited from $M$,\footnote{When $g$ is not flat, there are off-diagonal contributions to $G$. We use notation as if $g$ is flat to avoid carrying Christoffel symbols around.} symplectic structure $f$, and thus compatible almost-complex structure $K$:
\begin{equation}
    G = g_{ij} \dd q^i \dd q^j + g^{ij} \dd p_i \dd p_j\,,
    \quad
    f = \dd p_i \wedge \dd q^i\,,
    \quad
    K^A_B := G^{AC}f_{CB} 
    \,.
\end{equation}
The space of field configurations $\mathrm{Maps}(\Sigma,T^*M) = T^*\!\calM$ also inherits a Riemannian metric, symplectic structure, and almost-complex structure in the same way as before. We will call $N := T^*M$ and $\calN := T^*\!\calM$.

In a first-order formalism, the action necessarily becomes complex
\begin{equation}
    S[q,p] = \int \dd x\, \left(-i p_i \partial_x q^i + H(p,q) \right)\,,
\end{equation}
where $H(p,q) = \tfrac{1}{2} p^2 + V(q)$. Switching to the combined coordinates $X^A$ on $N$, the action is
\begin{equation}
    S[X] = \int \dd x\, \left(-ib_A\partial_x X^A + H(X)\right)\,,
\end{equation}
where $b = p_i \dd q^i$ is the symplectic potential for $f$. The corresponding Langevin equation with $2n$ real white noise variables is
\begin{align}\label{eq:LangevinX}
    \partial_t X^A(x,t) 
        &= - \frac{1}{2}G^{AB} \frac{\delta S}{\delta X^B} + \eta^A\\
        &= \frac{i}{2} K^A_B \partial_x X^B(x,t) - \frac{1}{2}G^{AB}\partial_B H(X)  + \eta^A(x,t)\,.
\end{align}
We see something curious: while the second-order quantum mechanics action gave entirely real Langevin evolution, the first-order presentation of the same theory gives a complex Langevin equation which immediately tries to drive $p$ off the real manifold $N$. Note: this is a consequence of the $i$ in the complex action, not the almost-complex structure $K^{A}_B$ on $N$. The origin of this $i$ can be traced back to having a Euclidean action $S$ which Wick rotates from a real-valued action $S_L$ of real fields in Lorentzian signature.

Since the first-order Langevin dynamics intrinsically forces us into a complex space, we may as well have considered a complexified manifold $N \hookrightarrow N^{\bbC}$ to begin with. We considered similar setups in Section \ref{sec:NormalizabilityAnalyticCont}, but now pay careful attention to the extension of all of our data $(G, f, K)$ on $N$. As a manifold,
\begin{equation}
    N^{\bbC} := (T^*M)^{\bbC} = T^{*}(M^{\bbC})
\end{equation}
is $4n$ real-dimensional with real coordinates $(X^A,Y^A) =: W^\alpha$. It has an integrable complex structure $J$ which can be used to define $J$-holomorphic and $J$-antiholomorphic coordinates
\begin{equation}
    Z^A := X^A + i Y^A\,,\qquad
    \overline{Z}^A := X^A - i Y^A\,.
\end{equation}
In terms of the natural holomorphically extended coordinates from $T^*(M^{\bbC})$, we could write $Z^A = (z^i, \pi_j)$. To be a complexification of $N$, we should also have an involution $\tau: N^{\bbC} \to N^{\bbC}$ that is $J$-antiholomorphic $\tau_* J = - J \tau_*$ and has the original $N$ as its fixed-point locus. And we do: $N$ is the subspace where all $Y^A = 0$.

A naive extension of the Langevin equation to $N^{\bbC}$ could involve simply holomorphically continuing every piece of data. For example, we could generalize our compatible triple on $N$ by
\begin{equation}
    G^{\bbC} 
        := G_{AB}^{\bbC}(Z) \dd Z^A \dd Z^B\,,
    \quad
    f^{\bbC} 
        := \dd \pi_i \wedge \dd z^i\,,
    \quad
    (K^{\bbC})^{A}_{B} := (G^{\bbC})^{AC}f^{\bbC}_{CB}\,,
\end{equation}
and the original action to
\begin{equation}
    S^{\bbC}[Z] = \int \dd x\,(-ib_A^{\bbC} \partial_x Z^A + H^{\bbC}(Z))\,.
\end{equation}
This gives a direct analytic continuation of the Langevin equation \eqref{eq:LangevinX} to
\begin{align}
    \partial_t Z^A(x,t) 
        &= -\frac{1}{2}(G^{\bbC})^{AB} \frac{\delta S^{\bbC}}{\delta Z^B} + (\eta^{\bbC})^A \label{eq:LangevinZ}\\
        &= \frac{i}{2} (K^{\bbC})^A_B \partial_x Z^B(x,t) - \frac{1}{2}(G^{\bbC})^{AB}\partial_B H^{\bbC}(Z)  + (\eta^{\bbC})^A(x,t)\,,
\end{align}
where $\eta^{\bbC} = \eta_R + i \eta_I$. A priori, these data define a generalized flow inside $N^{\bbC}$. These flows are still not tangent to the real manifold $N$, but they are at least all contained within $N^{\bbC}$.

An illuminating issue appears if we consider our noise correlations more carefully. In the original real Langevin equation, we had
\begin{equation}\label{eq:noise2pt}
    \expval*{\eta^A(t,x)\eta^B(t',x')}_{\mathrm{st}} = \lambda G^{AB} \delta(t-t')\delta(x-x')\,,
\end{equation}
we also could equivalently write a positive probability distribution:
\begin{equation}\label{eq:NoiseProbability}
    \exp({-\frac{1}{2\lambda}\int \dd t\, \dd x\,  G_{AB} \eta^A \eta^B})\,.
\end{equation}
This last expression was particularly useful for obtaining a path integral interpretation of stochastic correlators where the metric $G_{AB}$ weighting the Gaussian noise was the same as the metric in the gradient drift $G^{AB}\partial_B S$ in our SUSY map. Naively analytically continuing either equation leads to issues. Suppose we consider the naive analytic continuation of \eqref{eq:noise2pt}, then $G^{\bbC}$ specifies the noise correlations $\expval*{\eta^{\bbC} \eta^{\bbC}}_{\mathrm{st}}$ and $\expval*{\overline\eta^{\bbC} \overline\eta^{\bbC}}_{\mathrm{st}}$ by conjugation, but not the cross-correlations $\expval*{\eta^{\bbC}\overline{\eta}^{\bbC}}_{\mathrm{st}}$. In components, we only get the constraints
\begin{align}
    \expval*{\eta^A_R \eta^B_R}_{\mathrm{st}} - 
    \expval*{\eta^A_I \eta^B_I}_{\mathrm{st}} 
        &= \Re(\lambda(G^{\bbC})^{AB})\delta(t-t')\delta(x-x')\,,\\
    \expval*{\eta^A_R \eta^B_I}_{\mathrm{st}} + 
    \expval*{\eta^A_I \eta^B_R}_{\mathrm{st}} 
        &= \Im(\lambda(G^{\bbC})^{AB})\delta(t-t')\delta(x-x')\,,
\end{align}
leaving the problem underspecified. Instead, we could try to continue the much stronger \eqref{eq:NoiseProbability} directly, by complex linearity, this will specify all noise correlation functions. However, even if we just restrict our attention to $N$, the metric $G$ expands on $TN^{\bbC}|_N = TN \oplus JTN$ as
\begin{equation}\label{eq:GConN}
    G^{\bbC}|_{N} = G_{AB}(X) (\dd X^A \dd X^B - \dd Y^A \dd Y^B) + i G_{AB}(X) (\dd X^A \dd Y^B + \dd Y^A \dd X^B)\,.
\end{equation}
This means that
\begin{equation}
    G^{\bbC}(\eta,\eta)|_N = G(\eta_R, \eta_R) + 2i G(\eta_R, \eta_I) - G(\eta_I, \eta_I)\,.
\end{equation}
Thus the real noise fields $\eta_I$ are added to the path integral with divergent contributions and/or inappropriately signed 2-point functions.\footnote{The Parisi-Klauder complex Langevin prescription takes real noise $\eta_I = 0$ \cite{Parisi:1983mgm, Klauder:1983zm}. This completely specifies positive noise 2-point functions, but then $G^{\bbC}$ does not appear in \eqref{eq:NoiseProbability} and thus in the SUSY map. More general complex-noise prescriptions are possible \cite{Aarts:2009uq}.}


This last point highlights a workable strategy: one thing that would guarantee good positive noise correlations is to pick a positive real metric $(G_A)^{\alpha\beta}$ on $N^{\bbC}$, viewed as a $4n$-dimensional real manifold. Then, the relations between \eqref{eq:noise2pt} and \eqref{eq:NoiseProbability} would follow immediately. Of course, we also want the metric $G_A$ to agree with $G$ along $N$. In a best case scenario, we would love to have something like
\begin{equation}\label{eq:desirableG}
    G_A\vert_N \sim G_{AB}(X) (\dd X^A \dd X^B + \dd Y^A \dd Y^B)\,.
\end{equation}
As we will see, families of such $G_A$ are neatly determined by picking a different generalization of our $(G,f,K)$ triple.

First up is the symplectic structure. Following \cite{Witten:2010zr}, we equip $N^{\bbC}$ with a complex symplectic structure by choosing a closed non-degenerate holomorphic $\Omega \in \Omega^{(2,0)}(N^{\bbC})$. In particular, we will demand that $\Omega$ satisfies
\begin{equation}
    \Im\Omega|_N = f\,,\qquad
    \tau^*(\Omega) = -\overline{\Omega}\,.
\end{equation}
If we split
\begin{equation}
    \Omega =: \omega + i f\,,
\end{equation}
then $\omega|_{N} = 0$, so $N$ is Lagrangian with respect to $\omega$ and symplectic with respect to $f$. Moreover, this means that $J$ turns $\omega$ into $f$, $\omega(U,JV) = - f(U,V)$ for any vectors $U,V \in TN^{\bbC}$. Locally, we can write $\Omega = \dd\Lambda$ and $\Lambda := c + i b$. 

Now we can pick any almost-complex structure $I$ on $N^{\bbC}$ so that $\omega$ is type $(1,1)$ with respect to $I$ and positive, i.e. 
\begin{equation}\label{eq:conditionsOnI}
    \omega(IU, IV) = \omega(U,V)\,,
    \quad
    -\omega(U,IU) > 0\,.
\end{equation}
As noted in \cite{Witten:2010zr}, this means that $I$ must necessarily be different from $J$, because $\omega$ is type $(2,0) \oplus (0,2)$ with respect to $J$. Using this real symplectic form $\omega$ and almost-complex structure $I$ on $N^{\bbC}$, this allows us to equip $N^{\bbC}$ with a compatible positive metric
\begin{equation}\label{eq:GADefn}
    G_A(U,V) := -\omega(U,IV)\,.
\end{equation}
Thus, a good $G_A$ is determined once we have specified an $\omega$ and $I$.

We can do better though. While any such $G_A$ is sufficient for defining our noise correlations, it would be nice if $G_A$ agreed with $G$ for vectors in $TN$. In particular, we will demand, for vectors $U,V\in TN$, that
\begin{equation}\label{eq:conditionsOnG}
    G_A(U,V) = G(U,V)\,,\quad
    G_A(U,JV) = 0\,.
\end{equation}
The first condition is self-explanatory. The second condition says that, whatever $G_A$ may be, we would like it if the $J$-imaginary directions to $N$ are still perpendicular to $N$ for our choice of $G_A$, i.e., $JTN = (TN)^{\perp,G_A}$. This maintains the intuition that the $i$ in the original Langevin equation was driving us off of $N$ (into a perpendicular direction). Now, since $G_A$ was defined by the compatibility condition \eqref{eq:GADefn}, we also have $G_A(U,IV) = 0$ for any $U,V \in TN$ using the fact that $N$ was Lagrangian for $\omega$. These demands mean that
\begin{equation}
    I(TN) = JTN\,.
\end{equation}
This means that for any $U\in TN$ that $IU = JV$ for some $V\in TN$, the difference between $I$ and $J$ on $TN$ can be encoded in some map $A:TN \to TN$, i.e. $I = JA$. But we can immediately find such an $A$; for any $U,V\in TN$
\begin{equation}
    G(U,V) = G_A(U,V) = -\omega(U,IV) = -\omega(U,JAV) = f(U,AV) = G(U,KAV)\,,
\end{equation}
so $A = -K$. Consequently, we see that these two conditions imply an almost-hyper-K\"ahler looking condition between the three complex structures along $N$:
\begin{equation}\label{eq:almostKahler}
    I|_{TN} = -JK\,.
\end{equation}
See also the end of Section 2.7 of \cite{Witten:2010zr}. If possible, we would like to use this to define the extension of $K$ to all of $N^{\bbC}$.

If we set $H = 0$, with this continuation, the original Langevin equation \eqref{eq:LangevinX} is now effectively repackaged by the 4n real flow equations on the real variables $W^\alpha = (X^A, Y^A)$:
\begin{equation}
    \partial_t W^\alpha(x,t) = -\frac{1}{2}I^\alpha_\beta \partial_x W^{\beta}(x,t)  + \eta^\alpha(x,t)\,,
\end{equation}
and the noise correlations are defined by the positive real noise correlations:
\begin{equation}
    \expval*{\eta^\alpha(x,t) \eta^\beta(x',t')} = \lambda G_A^{\alpha\beta}\delta(t-t')\delta(x-x')\,.
\end{equation}
Since $I^{\alpha}_{\beta} = (G_A)^{\alpha\gamma} \omega_{\gamma\beta}$, these are the same flow equations occurring in the $A$-model with target $N^{\bbC}$ \cite{Witten:2010zr} (although, some relative sign changes are probably in order since we use stable Morse basins and $t \in [0,\infty)$). This suggests a connection between the stochastic quantization on the complexified cotangent bundle and the A-model. The reader may note that we set $H=0$. In principle, there is no difficulty in carrying $H$ along, but we would have to decide what to do with the corresponding term in the Morse flow equation. One choice is to extend the real Langevin equation by $- \frac{1}{2}G_A^{\alpha\beta}\partial_\beta \mathrm{Re}(H^{\bbC}(W))$, which is what naturally appears in \cite{Witten:2010zr}. We opt to drop this term since we do not know how to generically add a Hamiltonian to the $A$-model description of Lefschetz thimbles, except in very special cases like moment maps.

\paragraph{An Actual Example.} Consider our discussion in the previous section with $M=\bbR^{n}$, metric $g_{ij} = \delta_{ij}$ and $H=0$. Then $N=\bbR^{2n}$ with global coordinates $X^A = (q^i, p_j)$, and triple
\begin{equation}
    G 
        = 
    \begin{pmatrix}
        1_n & 0\\
        0 & 1_n
    \end{pmatrix}\,,\quad 
    f 
        = 
    \begin{pmatrix}
        0 & -1_n\\
        1_n & 0
    \end{pmatrix}\,,\quad 
    K
        = 
    \begin{pmatrix}
        0 & -1_n\\
        1_n & 0
    \end{pmatrix}\,.\label{eq:GfK}
\end{equation}
The original Langevin equation is just
\begin{equation}
    \partial_t X(x,t) = \frac{i}{2}K\partial_x X(x,t) + \eta(x,t)
\end{equation}
where $\expval*{\eta^A(x,t)\eta^B(x',t')}_{\mathrm{st}} = \lambda\delta^{AB}\delta(t-t')\delta(x-x')$.

Now we complexify $N^{\bbC} = \bbC^{2n} \cong \bbR^{4n}$. We write our underlying $4n$ real coordinates as
\begin{equation}
    W^\alpha = (X^A; Y^A) = (q^i,p_j;\tilde{q}^i,\tilde{p}_j)\,.
\end{equation}
where $z^i = q^i + i \tilde{q}^i$ and $\pi_j = p_j + i \tilde{p}_j$ etc. as before. In this basis, the complex structure $J$ is
\begin{equation}
    J = \begin{pmatrix}
        0 & -1_{2n}\\
        1_{2n} & 0
    \end{pmatrix}\,.
\end{equation}

The naive analytic continuation of the $(G,f,K)$ triple to $(G^{\bbC}, f^{\bbC}, K^{\bbC})$ has the exact same matrix representations as \eqref{eq:GfK}, where the $2n \times 2n$ matrices are now understood to be acting on the complex holomorphic coordinates $Z^A = (z^i, \pi_j)$. The Langevin equation becomes
\begin{equation}
    \partial_t Z(x,t) = \frac{i}{2} K^{\bbC} \partial_x Z(x,t) + \eta^{\bbC}(x,t)\,.
\end{equation}
In terms of the real coordinates $(X, Y)$, this becomes
\begin{equation}
    \partial_t X(x,t) = -\frac{1}{2} K\partial_x Y(x,t) + \eta_R(x,t)\,,
    \quad
    \partial_t Y(x,t) = \frac{1}{2}K \partial_x X(x,t) + \eta_I(x,t)\,.
\end{equation}
As before, to study correlation functions of $X$ and $Y$'s, we would need to specify the cross-correlations of the $\eta$-variables. A direct continuation of the noise two-point function only gives constraints $\expval*{\eta_R \eta_R}_{\mathrm{st}} - \expval*{\eta_I \eta_I}_{\mathrm{st}} = \lambda \delta$ and $\expval*{\eta_I \eta_R}_{\mathrm{st}} + \expval*{\eta_R \eta_I}_{\mathrm{st}} = 0$, so additional choices must be made. For example, in \cite{Aarts:2009uq} they choose
\begin{equation}
    \begin{pmatrix}
    \expval*{\eta_R \eta_R}_{\mathrm{st}} 
        & \expval*{\eta_R \eta_I}_{\mathrm{st}}\\
    \expval*{\eta_I \eta_R}_{\mathrm{st}}
        & \expval*{\eta_I \eta_I}_{\mathrm{st}}
    \end{pmatrix}
    = \lambda \begin{pmatrix}
        N_R & 0\\
        0 & N_I
    \end{pmatrix} \delta\,,\quad N_R = N_I + 1\,,\quad N_I \geq 0\,.
\end{equation}
Parisi and Klauder choose $N_I = 0$ \cite{Parisi:1983mgm, Klauder:1983zm}. The corresponding noise measure is
\begin{equation}
    \exp(-\frac{1}{2\lambda}\int \dd t\, \dd x \left(\frac{\eta_R^2}{N_R} + \frac{\eta_I^2}{N_I}\right))\quad \text{for $N_I >0$}\,.
\end{equation}
A direct continuation of the original noise measure gives non-positive correlation functions.

Now we consider the $A$-model continuation. Here we have many choices of $\Omega$ and $I$. The simplest choice of $\Omega$ compatible with our demands is actually just $\Omega = i \dd \pi_i \wedge \dd z^i = i f^{\bbC}$. So now we must pick an almost-complex structure on $I$ that satisfies \eqref{eq:conditionsOnI} and such that the compatible metric $G_A$ satisfies \eqref{eq:conditionsOnG}. Rather than systematically studying all of the $Sp(4n,\bbR)$ matrices satisfying \eqref{eq:conditionsOnI}, we jump to \eqref{eq:almostKahler} which suggests that any good choice will look approximately hyper-K\"ahler on $TN \subset TN^{\bbC}$. The best way to satisfy that \textit{is} just to use the hyper-K\"ahler choice on $N^{\bbC} \cong \bbC^{2n}$, i.e. we extend the almost-complex structure $K$ on $N$ to $\hat{K}$ on all of $N^{\bbC}$, picking (written in terms of real coordinates $W^\alpha = (X^A; Y^A)$):
\begin{equation}
    I = 
    \begin{pmatrix}
        0 & -K\\
        -K & 0
    \end{pmatrix}\,,\quad
    J = 
    \begin{pmatrix}
        0 & -1_{2n}\\
        1_{2n} & 0
    \end{pmatrix}\,,\quad
    \hat{K} = 
    \begin{pmatrix}
        K & 0\\
        0 & -K
    \end{pmatrix}\,.
\end{equation}
Then $I^2 = J^2 = \hat{K}^2 = -1$ and they satisfy the defining relations of a hyper-K\"ahler triple $JI = \hat{K}$, $I\hat{K} = J$ and $\hat{K}J = I$ (with reversed cyclic ordering to usual). In real coordinates, the symplectic form $\Omega$ and metric $G_A$ are thus
\begin{equation}
    \Omega =
    \begin{pmatrix}
        i K & - K\\
        -K & -i K
    \end{pmatrix}\,,\quad
    G_A = 
    \begin{pmatrix}
        1_{2n} & 0\\
        0 & 1_{2n}
    \end{pmatrix}\,.
\end{equation}
The metric $G_A$ is of course the simplest construction we could have chosen to extend $G$ to the whole $4n$-dimensional manifold, as anticipated in \eqref{eq:desirableG}.

With this choice, the Langevin equation extends to all $4n$ real-dimensional coordinates on the manifold
\begin{equation}
    \partial_t W(x,t) = - \frac{1}{2}I \partial_x W(x,t) + \eta(x,t)\,.
\end{equation}
In terms of the $(X,Y)$, this becomes
\begin{equation}
    \partial_t X(x,t) = \frac{1}{2} K\partial_x Y(x,t) + \eta_R(x,t)\,,
    \quad
    \partial_t Y(x,t) = \frac{1}{2}K \partial_x X(x,t) + \eta_I(x,t)\,,
\end{equation}
with a crucial minus sign removed from the stochastic differential equations. Moreover, the noise correlations now satisfy $\expval*{\eta_R \eta_R} = \lambda \delta$ and $\expval*{\eta_I \eta_I} = \lambda\delta$ positive and independent. In fact, this shows exactly how the A-model continuation differs in terms of noise correlations: in the direct analytic continuation, we have $\expval*{\eta^{\bbC} \eta^{\bbC}}$ and no cross-correlations; but in the A-model prescription we have pure cross-correlation.


\subsection{Example: WZW and Chern-Simons Theories}\label{sec:WZWModel}
A curious example to consider, especially given the various bulk-boundary correspondences lurking, is the WZW model \cite{Ardonne:2003wa, Dijkgraaf:2009gr}. The WZW action describes a 2d conformal field theory of group-valued fields $g:\Sigma \to G$ on a (boundaryless, say) Riemann surface $\Sigma$, by combining a Principal Chiral Model $S_{\mathrm{PCM}}$ with a Wess-Zumino term $\Gamma$:
\begin{equation}
    S_{\mathrm{WZW}}[g] 
        := S_{\mathrm{PCM}}[g] - i \Gamma[g]\,,\label{eq:WZWAction}
\end{equation}
where
\begin{align}
    S_{\mathrm{PCM}}[g] 
        &:= -\frac{1}{8\pi} \int_{\Sigma} \dd^2 \sigma\, \sqrt{\rho} \rho^{ij} \Tr[g^{-1}\partial_i g g^{-1}\partial_j g]\,,\\
    \Gamma[g]
        &:= \frac{1}{12\pi}\int_B \dd^3 \sigma\, \epsilon^{ijk}\Tr[g^{-1}\partial_i g g^{-1}\partial_j g g^{-1} \partial_k g]\,,
\end{align}
$\rho$ is a metric on $\Sigma$, $B$ is any 3-manifold with $\partial B = \Sigma$ as a boundary, and we usually weight the action by a ``level,'' a positive integer $k$ \cite{Wess:1971yu, Witten:1983tw, Witten:1991mm}.

As in the previous section, the $i$ in the Euclidean action leads to some issues. If we try to directly apply the Langevin equation to this action, the complex action immediately drives the field $g$ off of $G$ and onto $G^{\bbC}$. This leaves us trying to extend the real Langevin equation to the complexified space of fields, which is generically ambiguous as before. We anticipate that an extension of the Langevin equation exists connecting it to complex Chern-Simons \cite{Witten:2010cx}; we leave this to future work. 

Suppose instead that all we want is to find a (2+1)d theory whose ground state is $\Psi_0[g] = e^{-S_{\mathrm{WZW}}[g]/2\lambda}$. This is easy: we just define \cite{Ardonne:2003wa}:
\begin{equation}\label{eq:EoMWZW}
    D^{\pm} := \mp \frac{\delta}{\delta g} + \frac{1}{2\lambda}\frac{\delta S_{\mathrm{WZW}}}{\delta g}\,,
\end{equation}
then the (2+1)d Hamiltonian $H = \tfrac{\lambda}{2} D^+ D^-$ will formally have $\Psi_0[g]$ as a ground state, and $D^{\pm}$ give the 2d Dyson-Schwinger equations for $g$ (after changing to the topological frame). We could then supersymmetrize the bulk and identify bulk supercharges with boundary BV operators. An issue with this approach is that we lose the interpretation of the partition function as a state overlap because of the $i$ in the action, i.e.,
\begin{equation}
    |\Psi_0[g]|^2 = e^{-S_{\mathrm{PCM}}[g]/\lambda}\,.
\end{equation}
A similar issue happens with the first-order action in the previous section, suggesting that a completely satisfactory treatment must again involve analytic continuation.

\paragraph{A Gauged WZW Approach.} Instead, if we want a (2+1)d Hamiltonian whose ground state overlap/norm is the WZW partition function, there is a solution. The action \eqref{eq:WZWAction} has a $(G_L \times G_R) / Z(G)$ symmetry, so we can couple the $G_R$ symmetry to a background connection $A$ \cite{Witten:1991mm}:
\begin{equation}
    S_{\mathrm{gWZW}}[g,A] := S_{\mathrm{WZW}}[g] + \frac{1}{2\pi} \int_{\Sigma} \dd^2z\, \Tr A_{\bar{z}} g^{-1} \partial_z g - \frac{1}{4\pi} \int_{\Sigma} \dd^2z \, \Tr A_{\bar{z}} A_z\,.
\end{equation}
Here we are using complex coordinates $z$ and $\bar{z}$ on $\Sigma$. The $G_R$ symmetry is anomalous, so the action $S_{\mathrm{gWZW}}[g,A]$ is not invariant under $G_R$ gauge transformations. However, this specific choice of counterterms leaves gauge variations $\delta S_{\mathrm{gWZW}}$ independent of $g$, thus we can safely define the level-$k$ WZW partition function in the presence of the background field:
\begin{equation}
    \Psi[A] := \int [Dg] \,e^{-k S_{\mathrm{gWZW}}[g,A]}\,.
\end{equation}

$\Psi[A]$ can also be understood as a state of a ($2+1$)-dimensional theory, obtained by canonically quantizing the space $\calA$ of 2d connections (on the trivial bundle over $\Sigma$, say) \cite{Axelrod:1989xt, Witten:1991mm}.\footnote{We warn the reader that the bulk Chern-Simons theory discussion here is using Lorentzian/real-time and that the canonical quantization involves various $i$'s. We will proceed with a Lorentzian bulk for the remainder of this discussion since we are not going to connect directly to the Langevin equation anyway. This is also what happens in \cite{Ardonne:2003wa, Dijkgraaf:2009gr}. In Appendix \ref{app:SupersymmetryFormulas}, $H = -\tfrac{1}{2}\{Q_E,\bar{Q}_E\}$ generated Euclidean time evolution $-\partial_{t_E}$, so if we write $Q = Q_E$ and $Q^\dagger = - \bar{Q}_E$ then $H = \tfrac{1}{2}\{Q, Q^\dagger\}$ generates Lorentzian time evolution $i\partial_{t_L}$ and is properly Hermitian and positive semidefinite. Likewise, our conventions for this section will also be to put the physical states at fermion number $0$, rather than at top fermion number as was done in the earlier parts of the paper.} If we define gauge-covariant derivatives $D_i u = \partial_i u + [A_i,u]$ on fields, and covariant functional derivatives
\begin{equation}
    \frac{D}{D A_z} 
        := \frac{\delta}{\delta A_z} - \frac{k}{4\pi} A_{\bar{z}}\,,\quad
    \frac{D}{D A_{\bar{z}}} 
        := \frac{\delta}{\delta A_{\bar{z}}} + \frac{k}{4\pi} A_{z}\,,
\end{equation}
on the connections, then straightforward calculations show that $\Psi[A]$ satisfies
\begin{equation}
    \frac{D}{D A_z} \Psi[A] = 0\,,\quad
    \left(D_{\bar{z}}\frac{D}{D A_{\bar{z}}}-\frac{k}{2\pi}F_{\bar{z}z}\right)\Psi[A] = 0\,.\label{eq:DiffEqs}
\end{equation}
In too few words, the first means that $\Psi[A]$ is a holomorphic section of the $k$-th power of a certain complex line bundle $\calL^{\otimes k}$ over $\calA$, and the second means that $\Psi[A]$ is a gauge-invariant section of this line bundle, see \cite{Witten:1991mm} for more details. The upshot is that $\Psi[A]$ is a wavefunction of $G_k$ Chern-Simons theory, satisfying
\begin{equation}
    Z_{\mathrm{WZW}}(\Sigma) = \frac{1}{\mathrm{vol}(G)} \int [DA] \, |\Psi[A]|^2\,.
\end{equation}

This procedure hints at a general point. Correlation functions of $d$-dimensional theories -- or wavefunctions of ($d+1$)-dimensional theories -- can satisfy a number of Ward identities, Dyson-Schwinger equations, or other differential equations. For example, the WZW action gives the Dyson-Schwinger equations for $g$, and using \eqref{eq:EoMWZW} we can encode the equations of motion for $g$ into anti-fields which are the boundary values of superfields in a bulk theory with ground state $\Psi_0[g]$. Alternatively, we can couple the WZW theory to a background field $A$ and integrate over $g$, highlighting the theory's response to $G_R$ symmetry backgrounds. In this presentation, the closest we have to the equation of motion for $g$ is the second equation (it relies on using the $g$ equations of motion under the integral sign), which is a gauge-constraint for Chern-Simons theory. The upshot is that there is no unique bulk associated to the 2d theory, the choice of bulk depends on which differential equations (better, 2d complex) or ``space of conformal blocks'' we wish to use to build a (cohomological) ($2+1$)d field theory.

With this in mind, we can construct a non-topological ($2+1$)d supersymmetric field theory whose ground state satisfies both of the equations in \eqref{eq:DiffEqs} and squares to the WZW partition function. Let\footnote{$\calG$ is not exactly the second differential equation in \eqref{eq:DiffEqs}, it is the full gauge-variation generator on $\calA$. I.e., it is defined on the whole space of connections, not just the holomorphic connections relevant to a particular polarization.}
\begin{equation}
    \calE_a := \frac{D}{DA_z^a}\,,\qquad
    \calG_a := -i \left(D_z \!\calE + D_{\bar{z}}\frac{D}{DA_{\bar{z}}} - \frac{k}{2\pi}F_{\bar{z}z}\right)_a\,,
\end{equation}
then the conjugates are
\begin{equation}
    \calE_a^{\dagger} = -\frac{D}{DA_{\bar{z}}^a}\,,\qquad
    \calG_a^{\dagger} = \calG_a\,,
\end{equation}
and $\calE_a \!\Psi[A] = 0 = \calG_a\! \Psi[A]$. Since $\calG_a$ is the generator of gauge transformations,
\begin{equation}
    [\calG_a, \calG_b] = i {f_{ab}}^c \calG_c\,,\quad
    [\calG_a, \calE_b] = i {f_{ab}}^c \calE_c\,,\quad
    [\calE_a, \calE_b] = 0\,.
\end{equation}

Now introduce canonically anticommuting fermions $\chi^a$, $\chi^\dagger_a$, normalized so $\{\chi_a,\chi_b^\dagger\} = \delta_{ab}$, and let
\begin{equation}\label{eq:QEs}
    Q_{\calE} 
        := \int_{\Sigma} \dd^2z\, \chi_a^\dagger \calE^a\,,
    \qquad Q_{\calE}^\dagger 
        = \int_{\Sigma} \dd^2z\, \calE_a^\dagger\chi^a\,.
\end{equation}
It is easy to confirm that $Q_{\calE}^2 = 0 = (Q^\dagger_{\calE})^2$. Up to a normalization choice for $A$, these $Q_{\calE}$ and $Q_{\calE}^\dagger$ match the supercharges defined in (3.47) and (3.48) of \cite{Dijkgraaf:2009gr}. This can be used to define a K\"ahler SUSY quantum mechanics on the space of connections $\calA$, essentially giving an SQM interpretation of the geometric quantization of Chern-Simons theory \cite{Axelrod:1989xt} (although we should also consider the gauge constraints). Since the original states were holomorphic sections of the bundle $\calL^{\otimes k}$, i.e., $\bar{\partial}_{\calL^{\otimes k}}\Psi = 0$, the new fermions $\chi^\dagger$ effectively behave as anti-holomorphic 1-forms, thus our space of states is $\Omega^{0,p}(\calA, \calL^{\otimes k})$ with $Q_{\calE}$ the anti-holomorphic differential.\footnote{This supersymmetrization is a bit formal. Since $\calA$ is an affine space, we anticipate that the higher cohomologies $H_{\bar\partial}^{0,p}$ are zero, analogous to $\bbC^d$, and thus that there are no exciting BPS states. It is plausible that turning on further deformations/superpotentials to compute some relative cohomology or including higher holomorphic forms is more interesting, or we can change $\calA$ to an object with more interesting higher sheaf cohomology.}

We would also like to include the gauge symmetry constraints. To this end, we introduce canonically anticommuting fermions $c^a$ and $b_a$, normalized so $\{c^a, b_b\} = \delta^a_b$, and try to define BRST charges associated to the gauge-variation $\calG_a$. A putative BRST operator should be nilpotent and anti-commute with the $Q_{\calE}$'s. A set of BRST charges is \cite{Bomans:2025klo}:
\begin{equation}
        Q_{\calG} 
        := \int_{\Sigma} \dd^2z\, c^a \left(\calG_a + \calG_a^\chi + \tfrac{1}{2}\calG_a^{\mathrm{gh}}\right)\,,
    \qquad
        Q_{\calG}^\dagger 
        := \int_{\Sigma} \dd^2z\, \left(\calG_a + \calG_a^\chi + \tfrac{1}{2}\calG_a^{\mathrm{gh}}\right) b^a\,,
\end{equation}
where
\begin{equation}
    \calG_a^{\chi} 
        := - i {f_{ab}}^c \chi^{b\dagger} \chi_c\,,
    \qquad
    \calG_a^{\mathrm{gh}}
        := -i {f_{ab}}^c c^{b} b_c
\end{equation}
are the generators of the gauge-variation on the fermion and ghost sectors. We note that $Q_{\calG}$ is almost the lift of $\calG_a$ to the even larger phase-space of $(A, \chi^\dagger, \chi, c, b)$, but the ghost generator needs a factor of $1/2$ to compensate symmetry factors. One can check that $Q_{\calG}^2 = 0 = (Q_{\calG}^\dagger)^2$ and that all mixed $\calE$-$\calG$-supercharge anticommutators vanish.

Adding the BRST charges to the story is mathematically tensoring a Chevalley-Eilenberg complex for the Lie algebra of gauge transformations $\mathfrak{g}_\Sigma = \mathrm{Lie}(\mathrm{Maps}(\Sigma,G))$ to our space of anti-holomorphic forms. In other words, the proto-state space of our higher dimensional theory should be the bicomplex
\begin{equation}
    C^{p,q} = \Omega^{(0,p)}(\calA,\calL^{\otimes k}) \otimes C^q_{\mathrm{CE}}(\mathfrak{g}_{\Sigma})\,,\quad p,q \geq 0\,.
\end{equation}
The $U(1)_\chi$ and $U(1)_{\mathrm{gh}}$ symmetries grade the bicomplex with $Q_{\calE}$ and $Q_{\calG}$ carrying $+1$ under the respective charges. We can define a total differential on the bicomplex by
\begin{equation}
    Q := \alpha_{\calE} Q_{\calE} + \alpha_{\calG} Q_{\calG}\,.
\end{equation}
So long as the $\alpha_{\calE}$ and $\alpha_{\calG}$ are non-zero, the cohomologies of different $Q$ should be the same -- just with different representatives. The $\alpha$'s will help track the contributions of independent supercharges in the final Hamiltonian. Now we can define a manifestly supersymmetric Hamiltonian\footnote{The acronym stands for: supersymmetric covariant Carrollian strong-coupling Chern-Simons theory.}
\begin{equation}
    H_{\mathrm{SCCSCCS}} 
        := \frac{1}{2}\{Q, Q^\dagger\} 
        = \frac{e^2}{2}\{Q_{\calE},Q_{\calE}^\dagger\} + \frac{g^2}{2}\{Q_{\calG},Q_{\calG}^\dagger\}\,,
\end{equation}
where $e^2 := |\alpha_{\calE}|^2$ and $g^2 := |\alpha_{\calG}|^2$. It is highly non-obvious how the cohomology of $Q$ relates to the cohomologies of $Q_{\calE}$ and $Q_{\calG}$, and understanding this properly requires some spectral sequence calculations beyond the scope of this example. However, what is true is that $H^0(Q)$ will be exactly the gauge-invariant holomorphic sections of $\calL^{\otimes k}$, so that the BPS subsector of fermion-number $0$ states is the original space of Chern-Simons states.\footnote{Our fermion number conventions are opposite to previous sections for simplicity. I.e., we say $\chi_a$ and $b_a$ have $-1$ charge under $U(1)_{\chi}$ and $U(1)_{\mathrm{gh}}$, and annihilate the $0$-fermion state $\ket{0}$.}

The Lagrangian in the $0$-fermion sector (the rest can be obtained by supersymmetrization using $Q$) is interesting. We can re-write the $\calE_a$ in terms of the canonical momenta $\Pi^a_{z,\bar{z}} = -i \delta/\delta A_{z,\bar{z}}^a$, then
\begin{align}
    \calL_{\mathrm{SCCSCCS}}^{F=0} 
        &= \int_\Sigma \dd^2z\, \Tr (\Pi_z \dot{A}_z + \Pi_{\bar{z}} \dot{A}_{\bar{z}}) - H^{F=0}_{\mathrm{SCCSCCS}}\\
        &= \int_\Sigma \dd^2z\, \Tr (\!-i\calE\! \dot{A}_z + i \calE^\dagger\! \dot{A}_{\bar{z}} + i\frac{k}{4\pi}(A_z \dot{A}_{\bar{z}} - A_{\bar{z}}\dot{A}_z) - \frac{e^2}{2} \calE^\dagger\! \calE - \frac{g^2}{2}\calG^\dagger\! \calG\!)\,.
\end{align}
We can re-write the $-\tfrac{g^2}{2}\!\calG^2$ term by introducing an adjoint-valued Lagrange multiplier, which we call $A_0$ with foresight, as $A_0 \!\calG + \tfrac{1}{2g^2}A_0^2$. 
If we switch back to familiar real coordinates $\dd^2z = 2 \,\dd^2x$, $A_{z} = \tfrac{1}{2}(A_1 - i A_2)$, integrate out $\calE$ and $\calE^\dagger$, declare $F_{0i} := \partial_0A_i - \partial_i A_0 + [A_0,A_i]$, etc. then the action in the $F=0$ sector becomes:
\begin{equation}
    S_{\mathrm{SCCSCCS}}^{F=0} 
        = \int_{\Sigma \times I} \dd t\,\dd^2x \Tr(\frac{1}{e^2} F_{0i}F_{0i} + \frac{k}{4\pi}\epsilon^{\mu\nu\rho}(A_\mu\partial_\nu A_\rho + \tfrac{2}{3}A_\mu A_\nu A_\rho) + \frac{1}{g^2}A_0^2)\,.
\end{equation}
As in previous examples, this theory is non-relativistic. At finite $g^2$, this theory gives an energy penalty for violations of an emergent 3d Gauss' law, with an exact Gauss law constraint in the limit that $g^2 \to \infty$. 

In the limit $g^2\to\infty$, the theory reduces to a non-relativistic Yang-Mills Chern-Simons theory with only an electric field term, matching \cite{Dijkgraaf:2009gr, Ardonne:2003wa}. This gauge-invariant theory is a strong coupling limit of (2+1)d Yang-Mills Chern-Simons theory; likewise, the full gauge-invariant (2+1)d $D=1$ $\calN=2$ supersymmetric theory $\calL_{\mathrm{SCCSCCS}}$ can be obtained as a strong coupling limit of a (2+1)d $\calN=1$ Yang-Mills Chern-Simons theory \cite{Deser:1981wh, Dijkgraaf:2009gr}. It is curious to note that the relativistic (2+1)d $\calN=1$ SUSY algebra has two supercharges, but they are not nilpotent. This means that the nilpotence of $Q$ and $Q^\dagger$ only emerges in the strong coupling limit/\.In\"on\"u-Wigner contraction to the non-relativistic system. Moreover, the contraction is not a $c\to\infty$ non-relativistic contraction, but a Carrollian $c\to 0$ non-relativistic contraction (see Appendix B of \cite{Bagchi:2026emg}).

\appendix
\section{Background on \texorpdfstring{$\calN=2$}{N=2} \texorpdfstring{$D=1$}{D=1} Supersymmetry}\label{app:SupersymmetryFormulas}
The defining relations of the Lie superalgebra describing Euclidean $\calN=2$ $D=1$ supersymmetry are
\begin{equation}
    \{Q,Q\} = 0\,,\quad
    \{\Qb,\Qb\} = 0\,,\quad
    \{Q,\Qb\} = 2\sigma H\,,
\end{equation}
and a $U(1)$ $R$-symmetry group. $\sigma \neq 0$ is a parameter which differs by reference, we choose $\sigma = -1$ in the main text, but use chiral coordinates (explained below).
The superalgebra acts on $\bbR^{1|*2}$ by supertranslations, generated by the differential operators
\begin{equation}
    \calH = -\frac{\partial}{\partial t}\,,\quad
    \calQ = \frac{\partial}{\partial\theta} - \sigma \bar\theta \frac{\partial}{\partial t}\,,\quad
    \calQb = \frac{\partial}{\partial\bar{\theta}} - \sigma \theta \frac{\partial}{\partial t}\,.
\end{equation}
These act on coordinates by
\begin{equation}\label{eq:coordinateSUSY}
    t 
        \mapsto t - \sigma(\bar\epsilon\theta+\epsilon \bar\theta)\,,\quad
    \theta 
        \mapsto \theta + \epsilon\,,\quad
    \bar\theta' 
        \mapsto \bar\theta + \bar\epsilon\,.
\end{equation}
We introduce the covariant derivatives
\begin{equation}
    D = \frac{\partial}{\partial\theta} + \sigma \bar\theta \frac{\partial}{\partial t}\,,\quad
    \Db = \frac{\partial}{\partial\bar{\theta}} + \sigma \theta \frac{\partial}{\partial t}\,,
\end{equation}
which satisfy
\begin{equation}
    \{D,\Db\} = 2\sigma\frac{\partial}{\partial t}\,, 
\end{equation}
and
\begin{equation}
    \{D,\calQ\} = \{D,\calQb\} = \{\Db,\calQ\} = \{\Db,\calQb\} = 0\,.
\end{equation}

Consider a superfield $F(t|\theta,\bar\theta)$, we can expand it in components:
\begin{equation}\label{eq:Superfield}
    F(t|\theta,\bar\theta) := f_{00}(t) + \theta f_{10}(t) + \bar\theta f_{01}(t) + \theta \bar\theta f_{11}(t)\,.
\end{equation}
With these conventions
\begin{alignat}{3}
    [Q,F] &= - \calQ F 
        &&= -f_{10}(t) + \bar\theta (\sigma f_{00}'(t)-f_{11}(t)) + \theta \bar\theta (-\sigma f_{10}'(t))\,,\\
    [\Qb,F] &= - \calQb F 
        &&= -f_{01}(t) + \theta (\sigma f_{00}'(t)+f_{11}(t)) + \theta \bar\theta (\sigma f_{01}'(t))\,.
\end{alignat}

In the main text, we naturally use ``chiral superspace coordinates.'' In particular, if we define
\begin{equation}\label{eq:chiSS}
    \tau = t + \sigma \theta \bar\theta\,,\quad 
    \xi = \theta\,,\quad
    \bar\xi = \bar\theta\,,
\end{equation}
then $\tau$ is invariant under the chiral transform (only dependent on $\epsilon$) in \eqref{eq:coordinateSUSY}. In the chiral coordinates
\begin{alignat}{3}\label{eq:chiSSVFs}
    \calH 
        &= -\partial_\tau\,,\quad
    &\calQ
        &= \partial_{\xi}\,,\quad
    &\calQb
        &= \partial_{\bar\xi}-2\sigma\xi\partial_\tau\,,\quad\\ 
    \hphantom{\calH} 
        &\hphantom{=-\partial_\tau\,,}\quad
    &D
        &= \partial_{\xi}+2\sigma\bar\xi \partial_\tau\,,\quad
    &\bar{D}
        &= \partial_{\bar\xi}\,.
\end{alignat}
In the main text, we rename $\tau \mapsto t$.

Our conventions for integrals are that
\begin{equation}
    \int [\dd\bar{\xi}\dd\xi] \, \xi \bar{\xi} = 1 \,.
\end{equation}
We define a one-sided superspace delta function
\begin{equation}
    \delta_\partial^{(1|2)}(t|\bar\xi,\xi) := \delta_\partial(t)\delta(\xi)\delta(\bar\xi)\,,
\end{equation}
where $\delta_\partial(t)$ is a standard one-sided $\delta$-function integrating to $1$ on $[0,T)$. If $\calM_b \subset \bbR^{1|*2}$ is any supermanifold defined by the embedding $-t+b\bar\xi\xi \leq 0$, then
\begin{equation}
    \int_{\calM_b} [\dd t|\dd\bar{\xi}\dd\xi] \delta^{(1|2)}_{\partial}(t|\bar\xi,\xi) F(t|\bar\xi,\xi) = F(0|0,0)\,.
\end{equation}

\acknowledgments I would like to thank Jacob Abajian, Kevin Costello, Kolya Dedushenko, Diego Delmastro, Rajeev Erramilli,  Pietro Ferrero, Davide Gaiotto, Simeon Hellerman, Alessio Miscioscia, Fedor Popov, Surya Raghavendran, Dalton Sakthivadivel, Adar Sharon, Emilio Trevisani, and Edward Witten for comments and discussions throughout the very long duration of this project. I especially need to thank Davide Gaiotto for important comments on the draft. This work was supported by the NSERC CGS-D and NSERC PDF programs.

This project was initiated in 2022. Key ideas were introduced, and writing performed, independently of LLMs. ChatGPT (various models up to and including GPT-6) was used for additional grammar, notation, and algebra consistency checks, as well as for finding citations. All results were verified by the author before publishing; all remaining errors are the author's responsibility.

\bibliographystyle{JHEP}
\bibliography{refs.bib}

\end{document}